\documentclass[11pt]{MAS}
\usepackage[T1]{fontenc}
\usepackage[utf8]{inputenc}
\usepackage{mathptmx}
\usepackage{textcomp}
\usepackage{amsmath}
\usepackage{caption}
\usepackage{subcaption}
\usepackage{colortbl}
\usepackage[super]{nth}
\usepackage{tikz}

\definecolor{amethyst}{rgb}{0.6, 0.4, 0.8}
\definecolor{Dist0}{HTML}{4284f5}
\definecolor{Dist1}{HTML}{0bb830}
\definecolor{Dist2}{HTML}{ffe600}
\definecolor{DistMISS}{HTML}{ff2700}
\definecolor{DistMiss}{HTML}{FF3838}

\VolumeNo{2}\IssueNo{2}\Year{2021}
\usepackage[authordate,bibencoding=auto,strict,backend=biber,natbib]{biblatex-chicago}
\title{Darts Analysis}

\addauthor{Ayham Makhamra}{A}{1}
\email{amakhamra@mail.roanoke.edu}
\addauthor{Yelyzaveta Satynska}{A}{1}
\email{ysatynska@mail.roanoke.edu}
\addauthor*{Michael Weselcouch}{A}{1}
\addlocation{1}{School of Health, Science, and Sustainability, Roanoke College, Salem, VA}
\email{weselcouch@roanoke.edu}

\abstract
{
In this paper we examine the effectiveness of five mathematical models used to predict the outcomes of amateur darts games.  These models not only predict the outcomes at the start of the game, but also update their estimations as the game score changes.  The models were trained and tested on a dataset consisting of games played by amateur players involving students, faculty, and staff at Roanoke College.  The five models are: the null model, which is based only on the live scores, a logistic regression model, a basic simulation model, a time-adjusted simulation model, and a new variation of the Massey model which updates based on the current score.  We evaluate these models using two approaches. First, we compare their Brier scores. Second, we conduct head-to-head comparisons in a betting game in which one model sets the betting odds while the other places bets.   In both cases, model performance is assessed not only at the start of the game but also at the start of each round.  Across both evaluation methods, the score-dependent Massey model performs the best. We conclude by illustrating how this score-dependent Massey model framework can be adapted to other competitive settings beyond darts.
}

\begin{document}

\maketitle

%%%%%%%%%%%%%%%%%%%%%%%%%%%%%%%%%%%%%%%%
%%%%%%%%%%%%%%%%%%%%%%%%%%%%%%%%%%%%%%%
\section{Introduction}
%\wthought{This is a rough draft, but I just wanted to get something on paper.  I'll edit it later.}

Every spring, the Roanoke College math department hosts a semester-long sports tournament called the Minton Invitational, named after Professor Emeritus and sports enthusiast, Roland Minton.  The sport that is played in the Minton Invitational changes from year to year, but there are certain things that have stayed consistent throughout the years.  The Minton Invitational is separated into two parts: the regular season and the postseason.  The games in the regular season do not follow a set schedule.  Players, who can be either students, faculty, or staff, create their own schedule and automatically qualify for the postseason once their first game is recorded. In the postseason, players can still schedule matches freely, but an elimination bracket is also introduced, creating a mix of self-scheduled and bracket-determined matchups. We will not be differentiating between self-scheduled and bracket-determined games in this article.%\lthought{Might be not important, but if we wanted to be precise, those that played <4 games do have get a rank - LS} 

To record the results of the games, a website was created that allows players to submit their scores anytime of the day and night. %\wthought{Maybe we should have a sentence highlighting Roland's contributions to sports analytics?}\lthought{Yes!}
The purpose of the Minton Invitational is to help build a sense of community between the students, faculty and staff, and to give students a hands-on experience with sports analytics as the datasets created from the Minton Invitational can be used for student research projects.
%\lthought{Consider something like: "...and to allow for the opportunity to conduct student research in sports analytics using the real-world data collected". The exisiting one sounds good too, but I thought it might come off as if all the pariticipating in the tournament students then conduct research}  
If you need help organizing a tournament at your school, do not hesitate to contact the third author. %\lthought{Love this! We could consider adding what exactly we could help with? E.g. providing the code for the website, tips for engaging the students into the game, methods to analyze the collected data, etc.}

The game played for the spring 2025 Minton Invitational was a variation of darts, called \emph{Darts 271}.  

\subsection{Darts 271}
\begin{figure}
    \centering
    \includegraphics[width=0.25\linewidth]{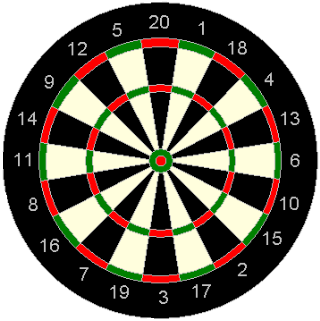}
    \caption{A traditional dartboard.}
    \label{fig:dart_board}
\end{figure}

Darts 271 is played with a traditional dartboard.  A dartboard has 20 numbered sectors, plus a bullseye (see Figure \ref{fig:dart_board}). Each numbered sector contains three sub-regions: the outer ring, the inner ring, and the areas in between. In Figure \ref{fig:dart_board}, the outer ring and inner ring are both colored  in red and green.  A dart that lands in the outer ring of a numbered sector scores double the number's value.  A dart that lands in the inner ring of the a numbered sector scores triple that number's value.  A dart that lands anywhere else in the numbered sector scores that number's value.  The outer bullseye (colored in green in Figure \ref{fig:dart_board}) is worth 25 points and the inner bullseye is worth 50 points.  Since most players competing in the Minton Invitational have never thrown a dart before, it was not uncommon for a dart to miss the dartboard completely.
In that case, the throw results in 0 points.

In a traditional darts game, players compete head-to-head by taking turns throwing three darts each to earn points.  In order to win, a player must finish with exactly 501 points with the last dart landing in the outer ring. 
Darts 271 has a lot of the same rules as traditional darts.  Like traditional darts, players compete head-to-head by taking turns throwing three darts to earn points.  Once both players throw their three darts, the round ends and the next round begins.  A major difference between traditional darts and Darts 271 is that, in Darts~271, the games are played to 271 points and players do not have to score exactly 271 points, they can surpass this value. 
Additionally, players do not need to end the game by landing in the outer ring.

In Darts 271, scores are updated after each round, meaning that the points are tallied after all six darts (three from each player) are thrown. Consequently, it is possible for a game to be tied with both players above the 271 point threshold.  In that case, rounds continue to be played until one of the players takes the lead.  Unlike innings in a baseball game or ends in a curling match,  
%\lthought{If we want to add to our bibliography, we could add a reference here to some paper explaining baseball and curling game rules} \wthought{IDK if that's necessary.}
the number of rounds in a Darts 271 game varies depending on the abilities of the players.  If both players have a hard time scoring points, the game can last many rounds.  Conversely, a game can have only a few rounds if even just one of the players consistently throws their darts into high value regions.  %\wthought{Maybe we can include a histogram of the number of rounds in each game?}

%\wthought{Again, I'm just getting words on the page.  Will edit to make better eventually.  }

\subsection{Modeling Framework and Evaluation}
The world of sports analytics has changed a lot in recent years due to the overwhelming presence of sports betting.  No longer is it the case that all bets must be placed before the games begin.  Many sports books now let you bet during the game and they update the odds based on a number of factors.  However, many of the classical sports ranking systems (Massey, Colley, Elo, etc.) are used to compare teams or competitors only at the start of games.  For an overview of the classical sports ranking methods, see the fantastic book ``Who's~\#1?" by \cite{meyer_langville_2010}.

In this article, we provide five models that estimate the probability of a player winning a Darts 271 game based off two factors: the players involved and the live score of the game.
The five models are: the null model which is based only on the live score, a logistic regression model, a basic simulation model, a time-adjusted simulation model, and a score-dependent variation of the Massey model.
%a live scores-dependent, linear algebra-based \lthought{Should we name this to "Massey-ratings-based"} model.  

We evaluate these models using two approaches. First, we compare their Brier scores. Second, we conduct head-to-head comparisons in a betting game in which one model sets the betting odds while the other places bets.  Further details on these evaluation methods are provided in Section \ref{sec: Data and Evaluation}. In both cases, model performance is assessed not only at the start of the game but also at the start of each round.

\subsection{Outline}
This paper is organized as follows: in Section \ref{sec: Data and Evaluation}, we discuss the dataset that we used, summary statistics calculated from the dataset, and the methods used to evaluate the models; in Section \ref{sec: Models}, we discuss the implementation and motivation behind each model; in Section \ref{sec: Results}, we discuss the performance of each of the models;
and Section \ref{Sect: Conclusion} contains our conclusions and directions for future work.

\section{Data and Evaluation}\label{sec: Data and Evaluation}

In this section we discuss our methods for collecting data, statistics that we calculated from the dataset, and a detailed description of the two methods used to evaluate the models.

\subsection{Data Collection}
For data collection, we have collaborated with the Roanoke College Information Technology department to develop a user-friendly website, where all Roanoke College (RC) affiliated public (students, faculty, and staff) can record their scores. The developed website requires the users to log in with their RC credentials, which reduces the risk of receiving spam or malicious data. Additionally, requiring participants to log in with their RC credentials allows us to be certain that the players' names stay consistent across the whole dataset since "Player 1" and "Player 2" fields are drop-down menus with the names of RC-affiliated public, and not free entries.

\begin{figure}[htbp]
  \centering
  \begin{subfigure}[t]{0.25328\textwidth}
    \centering
    \includegraphics[width=\linewidth]{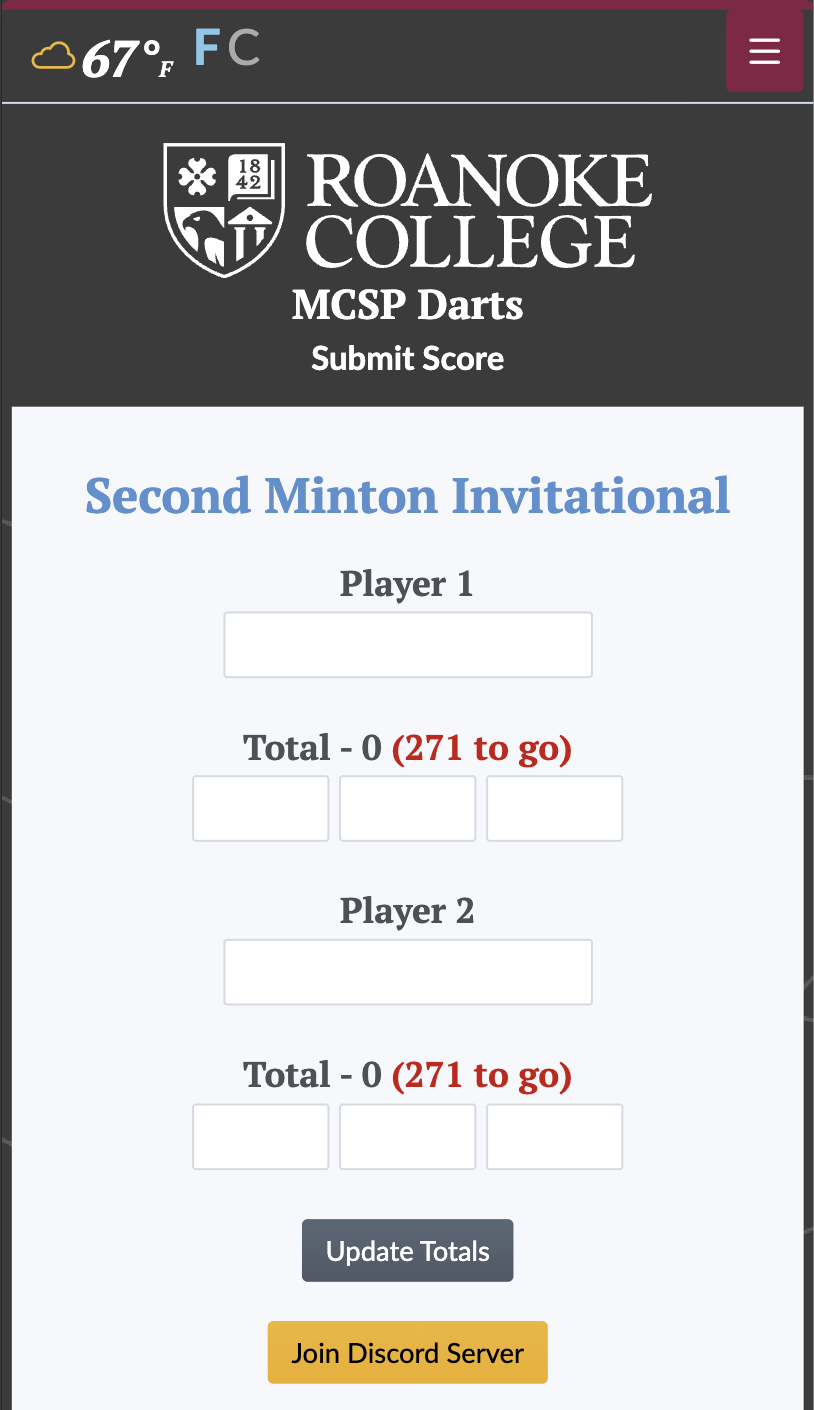}
    \caption{Mobile View}\label{fig:left}
  \end{subfigure}\hfill
  \begin{subfigure}[t]{0.73\textwidth}
    \centering
    \includegraphics[width=\linewidth]{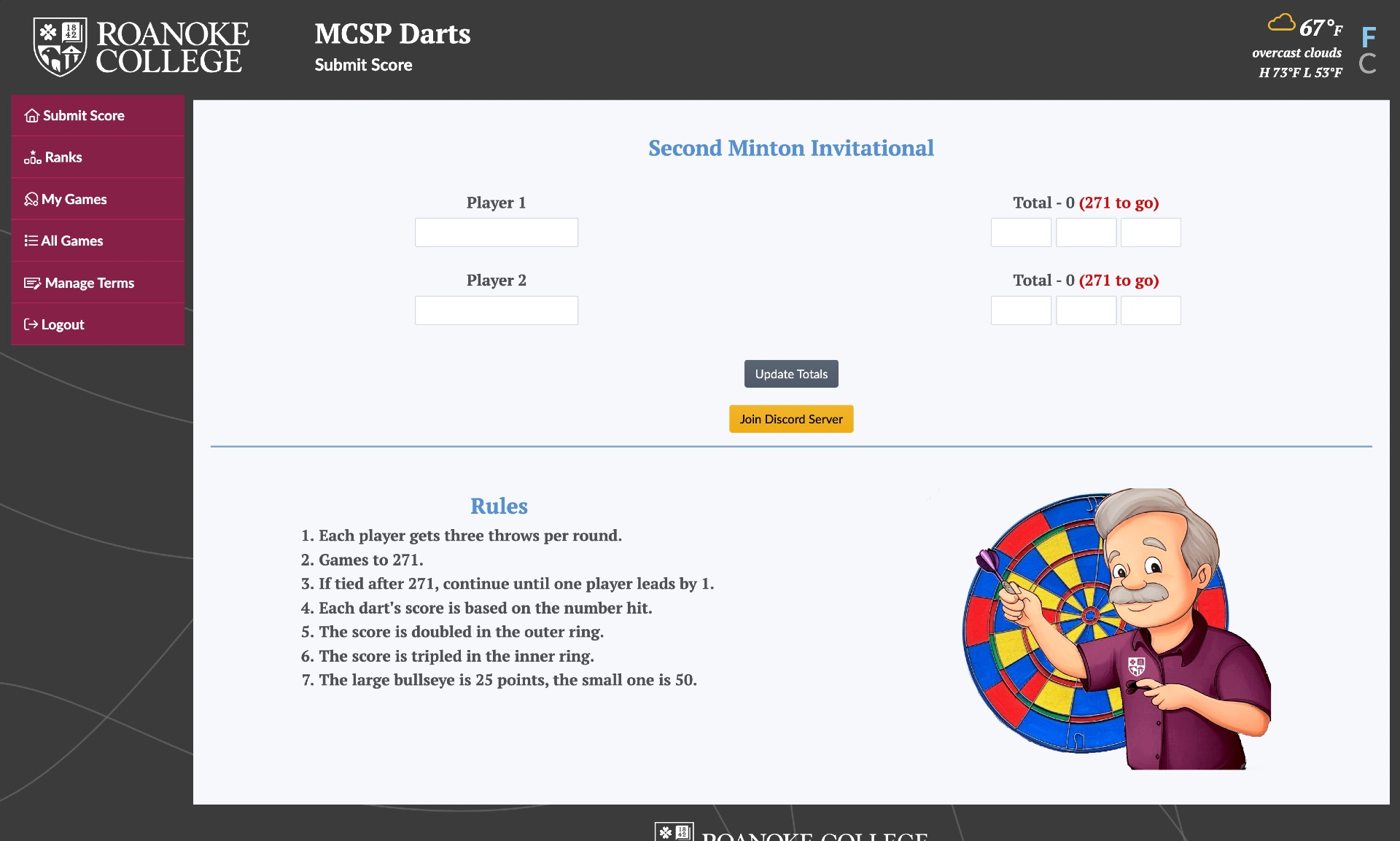}
    \caption{Desktop View}\label{fig:right}
  \end{subfigure}
  \caption{\emph{Darts 271} website for data collection.}
  \label{fig:website}
\end{figure}

The website is mobile-friendly as well as computer-friendly (see Figure~\ref{fig:website}), which allowed us to place a QR-code leading to the website beside the dartboard. This website served two purposes.  It allowed players to easily keep track of the score during the game, and it allowed us to have a detailed record of every throw recorded during the season. To use the website, a player simply logs in which automatically pre-populates their name in the "Player 1" field. Once the opponent's name is entered, the game is ready to start.

%\st{The website was designed to be player--friendly.  To use it, a player simply logs in which automatically pre-populates their name in the "Player 1" field. Once the opponent's name is entered, the game is ready to start.} \lthought{Trying to remove the repetition of ``-friendly" in this and the previous paragraph}

The six scores from the round are then entered and the player selects "Update Totals". This computes the current score and clears the values submitted from the previous round.  By clearing the values, the risk of a player accidentally submitting the same round twice is reduced. The website automatically determines the winner once a player is leading and either of the total scores becomes more than 271. At that point, no more rounds can be submitted.

We plan to update the website to display mid-game win probability estimation values using the models described in Section \ref{sec: Models}.

\subsection{Dataset Overview}\label{SubSect: Dataset Overview}

% The dataset used in this research included detailed records from the Minton Invitational starting from February \nth{6} until May \nth{2}. 
% We had 45918 individual throws over 1131 games played with 45 players participated in this tournament. Each row in this dataset holds important information including: Timestamp of the throw, players' id, players' name, game id, round number, score per throw (However, in this dataset we didn't record the multipliers 2x, 3x, instead we recorded the full score) for each player in that game. 
% This simple though rich dataset allowed us to later model each player performance and strategies over time. 

\begin{figure}[htbp]
  \centering
  \begin{subfigure}[t]{0.45\textwidth}
    \includegraphics[width=\linewidth]{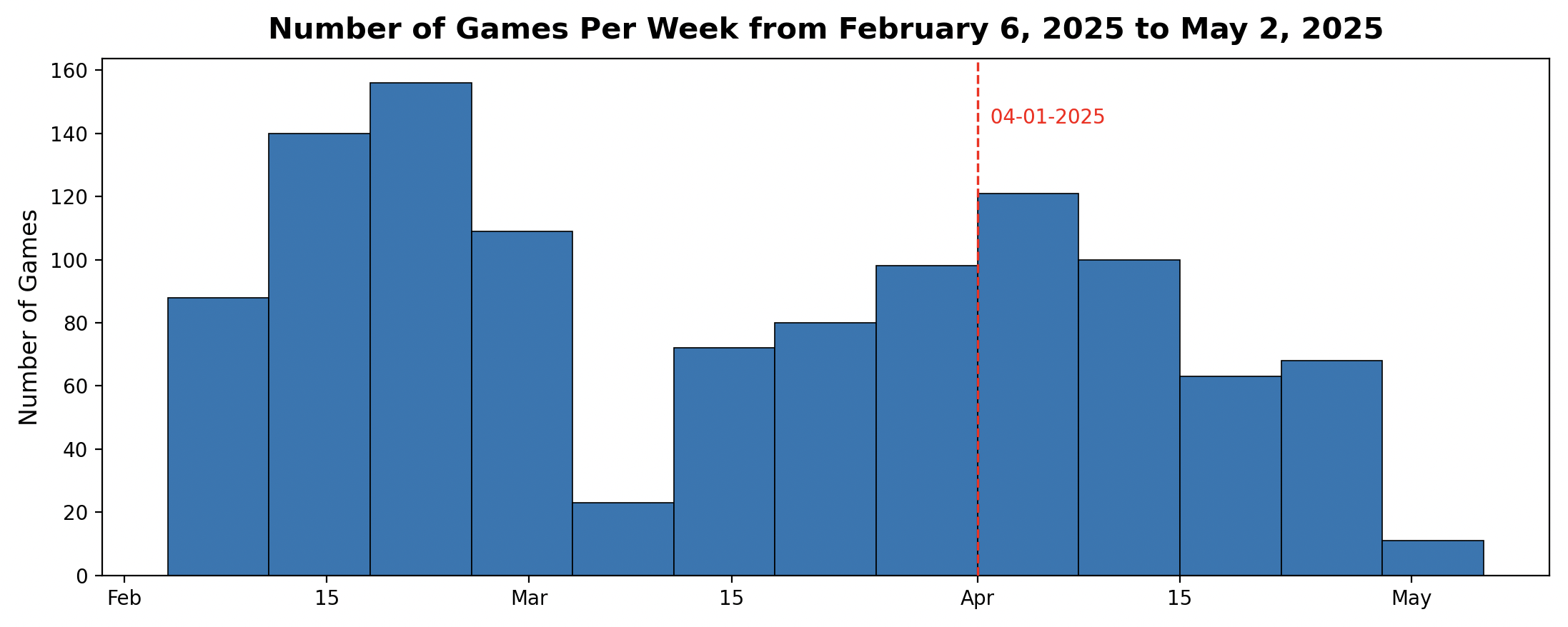}
        \caption{Weekly game counts throughout the season.}
    \label{fig:number_games_histogram}
  \end{subfigure}\hfill
  \begin{subfigure}[t]{0.45\textwidth}
    \includegraphics[width=\linewidth]{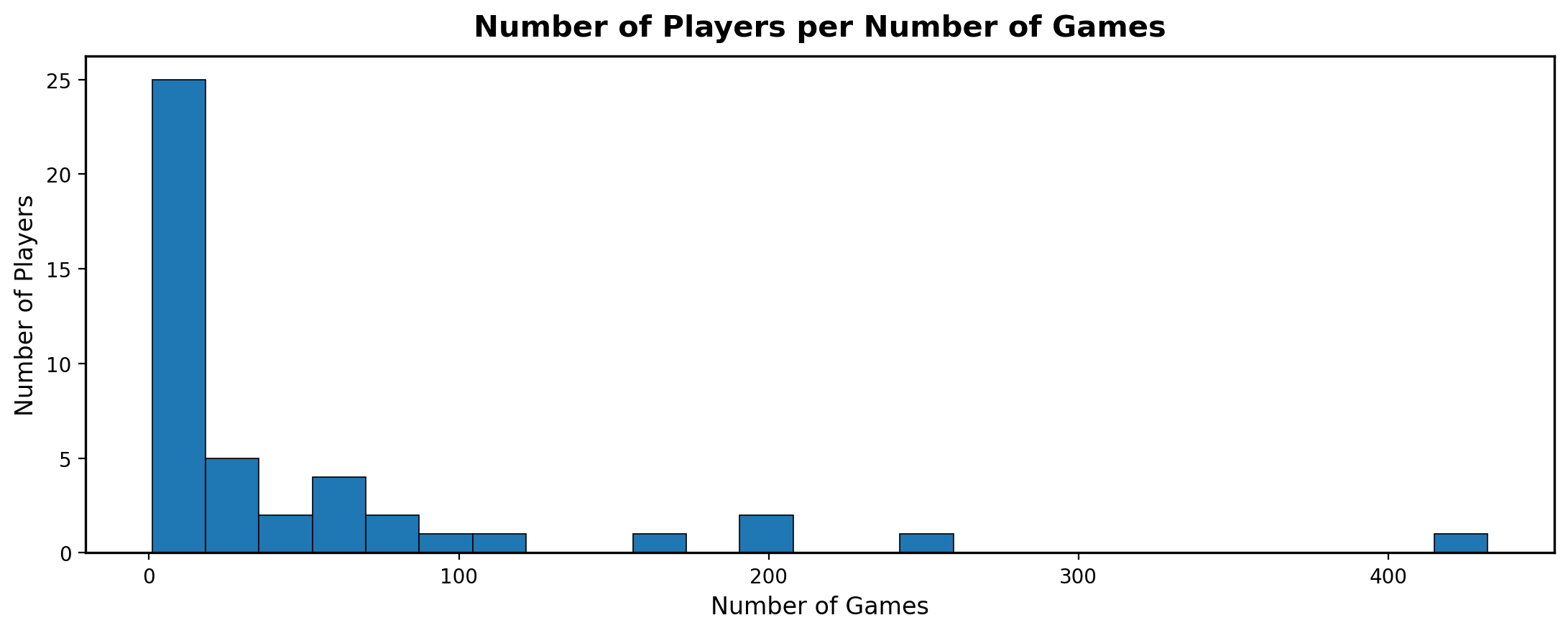}
    \caption{Number of games distribution for all players.}
    \label{fig:number_of_games_distribution}
  \end{subfigure}

\vspace{.75 cm}
  
  \begin{subfigure}[t]{0.45\textwidth}
  \includegraphics[width=\linewidth]{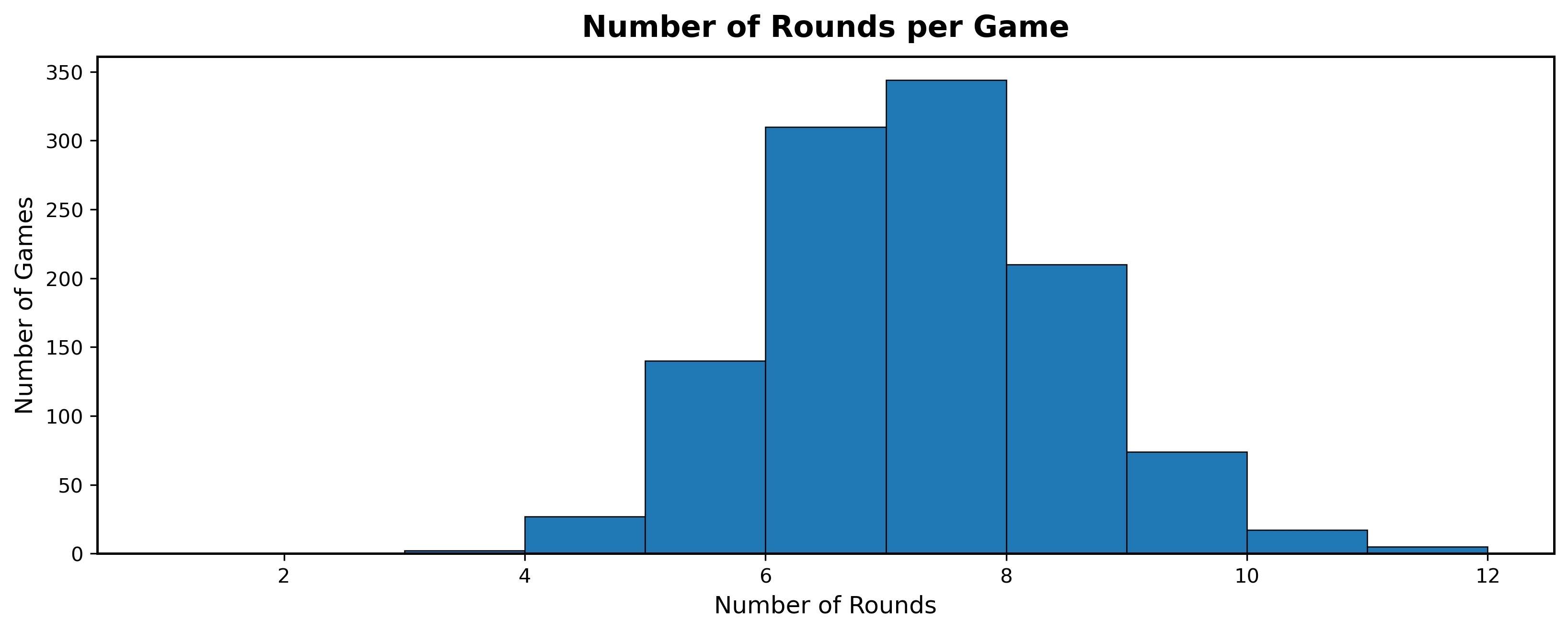}
    \caption{Number of rounds distribution for all games.}
    \label{fig:number_of_rounds_distribution}
  \end{subfigure}\hfill
  \begin{subfigure}[t]{0.45\textwidth}
  \includegraphics[width=\linewidth]{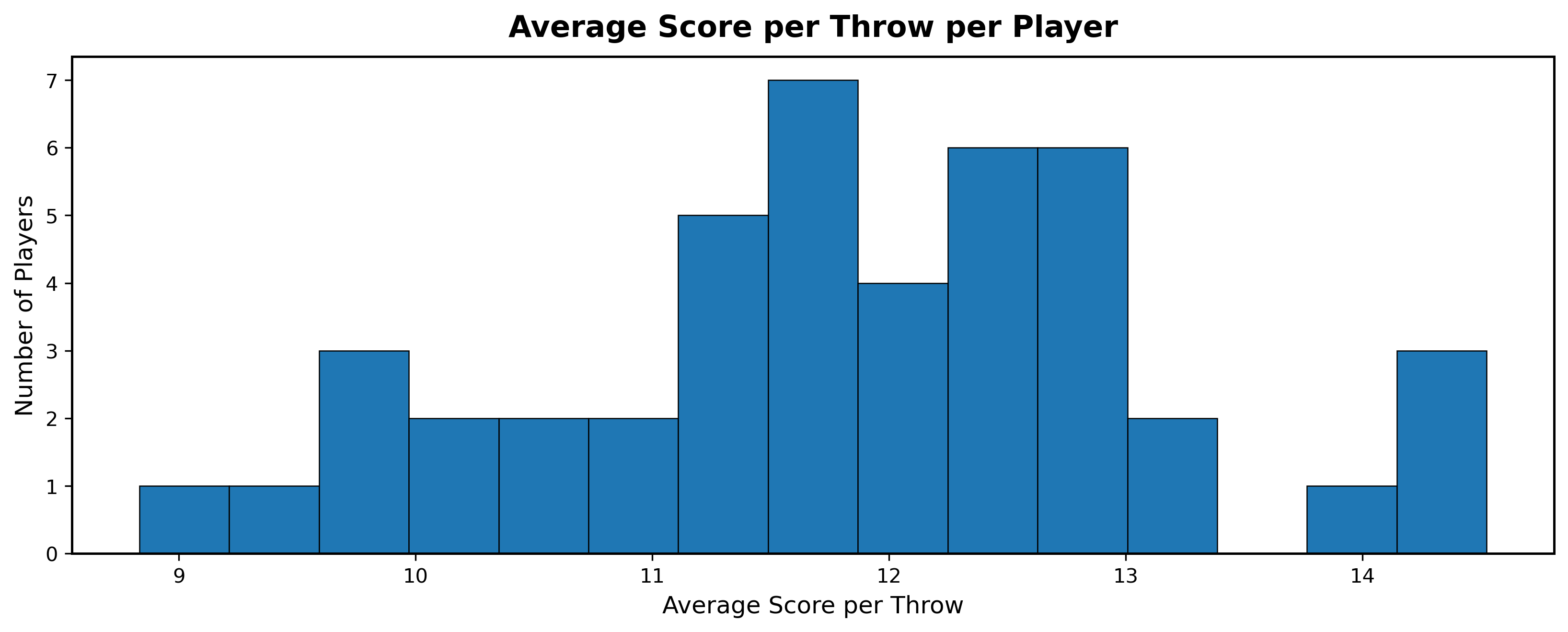}
    \caption{Average score distribution for all players.}
    \label{fig:avg_scores_distribution}
  \end{subfigure}\hfill
  \caption{Histograms describing key features of the dataset.}
  \label{fig:Summary Histograms}
\end{figure}

We collected data from February 6, 2025 to May 2, 2025 using the developed website, where the data up until April 1, 2025 constituted our training set, and after April 1, 2025, the testing set.  A total of 1,131 games were recorded (see Figure \ref{fig:number_games_histogram}) collectively consisting of 15,306 rounds spanning over 45,918 throws. In total, 45 players participated in the competition.  Each row in our dataset corresponds to one recorded throw.  The following information was collected for each throw recorded: the start time of the game, the player IDs of the thrower and their opponent, the game ID, the round number, and the points scored on the throw.

We note some of the limitations of our dataset. Firstly, the data does not allow us to determine which player threw first, which could have been of value if we were to predict whether the order of throwing matters. Additionally, we are not recording the information about the order of the three throws in each round. Finally, there are scores that can be obtained in multiple ways, for example a score of 14 can either be from the single 14 region or the double 7 region, which our data does not differentiate between. Because of these three limitations, the analysis was done under the assumption that all six throws from each round occur simultaneously, preventing us from being able to do mid-round analysis.

Each player managed their own schedule, choosing when and how often to compete throughout the season.
Figure \ref{fig:number_of_games_distribution} shows the distribution of games played per player throughout the season. Most players participated in fewer than $40$ games, while $7$ players played $100$ or more games over the course of the season.  Consequently, players who played more games are overrepresented in the dataset  which is an unavoidable limitation of the league’s self-scheduling format.

%\athought{Following this, Figure \ref{fig:number_of_rounds_distribution} we see distribution of the number of rounds per game. Most games consist of 6 to 9 rounds, with the peak at 7 rounds. Very few games extend beyond 10 rounds showing extended games are uncommon.}

Recall that a game of Darts 271 does not have a fixed number of rounds.  The game continues until a player leads with a score of or above 271.  Figure \ref{fig:number_of_rounds_distribution} shows the distribution of the number of rounds played in a game.  The average number of rounds in a game was $6.78$, and most games were completed within $5$ to $9$ rounds. While none of the models introduced in Section \ref{sec: Models} account for round numbers, incorporating features such as a player’s typical game length could potentially improve the predictive accuracy of the models.

%\athought{Figure \ref{fig:avg_scores_distribution} presents the distribution of average scores per player. The majority of players have an average score between 11 and 13 points per throw, with a few players achieving averages above 14. This disparity suggests the presence of a subset of players with significantly higher performance, potentially indicating greater skill or experience compared to the majority.}

Finally, Figure \ref{fig:avg_scores_distribution} shows the distribution of average points per throw for each player. Most players averaged between 11 and 13 points per throw, while a few exceeded 14 points. This suggests that the gap between the top players and the rest of the field is relatively small, only a few points per throw on average. As a result, even less-skilled players have a realistic chance of winning against any opponent.  This makes upsets relatively common and making accurate game predictions particularly challenging.

%\begin{figure}
%    \centering
%    \includegraphics[width=0.6\linewidth]{new_games_dist_hist.png}
%    \caption{Weekly game counts throughout the season.}
%    \label{fig:number_games_histogram}
%\end{figure}

%\begin{figure}
%    \centering
%    \includegraphics[width=0.6\linewidth]{Histogram_of_Number_of_Games.png}
%    \caption{Number of games distribution for all players.}
%    \label{fig:number_of_games_distribution}
%\end{figure}

%\begin{figure}
%    \centering
%    \includegraphics[width=0.6\linewidth]{Number_of_Rounds_per_Game.png}
%    \caption{Number of rounds distribution for all games.}
 %   \label{fig:number_of_rounds_distribution}
%\end{figure}

%\begin{figure}
%    \centering
%    \includegraphics[width=0.6\linewidth]{Distribution_of_Average_Scores_per_Player.png}
%    \caption{Average score distribution for all players.}
%    \label{fig:avg_scores_distribution}
%\end{figure}

% \begin{figure}[htbp]
%   \centering
%   \begin{subfigure}[b]{0.45\textwidth}
%     \centering
%     \includegraphics[width=\textwidth]{Distribution_of_Average_Scores_per_Player.png}
%     \caption{Average score distribution for all players}
%     \label{fig:avg_scores_distribution}
%   \end{subfigure}
%   \hfill
%   \begin{subfigure}[b]{0.45\textwidth}
%     \centering
%     \includegraphics[width=\textwidth]{Average_Score_vs._Player_Consistency.png}
%     \caption{Average score and Player consistency distribution}
%     \label{fig:avg_score_VS_player_consistency}
%   \end{subfigure}
%   \caption{Some dataset summary statistics}
%   \label{fig:combined}
% \end{figure}

\subsection{Brier Score}

The \emph{Brier score} of a model is a numerical value that can be used to evaluate the accuracy of the model.  The Brier score can be used when the potential outcomes of the variable are categorical and mutually exclusive.

Let $\pi = (\pi_1, \pi_2, \dots, \pi_n)$ be a model that forecasts the probability of an event, where $\pi_t$ is the predicted probability of the event occurring at instance $t$. If $\sigma_t \in \{0, 1\}$ is the actual outcome of the event at instance $t$ ($\sigma_t = 0$ means that the event did not occur, $\sigma_t = 1$ means that the event did occur), then, after $n$ predictions are made, the Brier score $B(\pi)$ of the model is given by
\[B(\pi) = \frac{1}{n}\sum_{t=1}^n(\pi_t - \sigma_t)^2.\]
That is to say, the Brier score measures the mean squared difference between the predicted probabilities and the actual outcomes.

The Brier score ranges from $0$ to $1$, with lower values indicating better-calibrated predictions. Note that a model can be perfectly calibrated yet still have a Brier score greater than $0$. For example, a model that assigns a $50\%$ probability to a fair coin landing heads will have a Brier score of $0.25$, even though its probabilities are exactly correct.  In general, if an event has a constant probability $p$, then the Brier score of a perfectly calibrated model will tend towards $p(1-p)$ as the number of predictions tends to infinity.

This statistic is commonly used to evaluate the accuracy of sports prediction models, since a typical sporting event provides many opportunities to test a model’s forecasts. Recently, \cite{clemens2022appletvprobabilities} published an article on \emph{FanGraphs}  in which he evaluated the accuracy of the ``controversial” Apple TV prediction model using Brier scores. Clemens found that the probabilities displayed on the Apple TV broadcasts of Major League Baseball games produced higher (worse) Brier scores than those from a simple, single-variable, baseline model he developed.  His analysis provided statistical evidence of the Apple TV model’s shortcomings. %\lthought{Haha, I like it but are we sure we can write this? I'm worried that it sounds more like an opinion than a fact.} \wthought{I say keep it.  The referee can always tell us to cut it.}

A modified version of the Brier score was used in a recent study by \cite{brown2025predictive} where the authors evaluated the accuracy of several linear algebra–based models that predicted the outcomes of lower-division English soccer matches. The modification accounted for the fact that a soccer match can end in one of three possible outcomes: a win, a loss, or a draw.  While we won't be using the three-outcome variation of the Brier score in this paper, future research may consider extending this work to account for multiple outcomes or perhaps to weigh predictions by their predicted size of victory. %\lthought{Should we elaborate on how this could be relevant to our work?} \wthought{I agree.  We should have one more sentence in this paragraph. What do you think of the one I added?}

%\lthought{I really like this example with the .25 Brier score being exactly correct. I think it would be cool if we could add another brief example from one of the sources you listed.} \lthought{I added a reference though it might be not in the proper format~\cite{clemens2022appletvprobabilities}}  \wthought{I want to include a citation to a Fangraphs article that computed the Brier score for Apple TV's baseball model just to include more context for possible Brier scores.  I'm not sure how to include it yet, but here's the article https://blogs.fangraphs.com/how-good-are-those-probabilities-on-the-apple-tv-broadcasts/.  Any idea how to cite a blog?  That might be a first.  Also, we can reference this paper too. https://janeway.uncpress.org/ms/article/1327/galley/2113/view/}

\subsection{Betting Game}\label{SubSect Betting Game}
Another way to compare two models is to imagine a game in which players gamble against each other using the models’ predictions. This setup is called a \emph{betting game}. To compare model $\alpha$ against model $\beta$, player A sets the odds of the event occurring at instance $t$ to $\alpha_t$. Player B then places a bet that the event will occur if $\beta_t > \alpha_t$, and that it will not occur if $\beta_t < \alpha_t$.  If both models assign the same probability, no bet is placed for that instance.  Once the outcome is determined, payout is assigned.  If the event occurs, the payout is $|\alpha_t-\beta_t|\cdot(1-\alpha_t)$ if player B bets correctly, and %$-|\alpha_t-\beta_t|\cdot\alpha_t$ 
$-|\alpha_t-\beta_t|\cdot(1-\alpha_t)$ if player B bets incorrectly.  Similarly, if the event does not occur, the payout is $|\alpha_t-\beta_t|\cdot \alpha_t$ if player B bets correctly, and %$-|\alpha_t-\beta_t|\cdot(1-\alpha_t)$ 
$-|\alpha_t-\beta_t|\cdot \alpha_t$ if player B bets incorrectly.  The payouts are shown in Table \ref{tab:payouts}.
%The payout for a correct bet is $|\alpha_t-\beta_t|\cdot\frac{1-\alpha_t}{\alpha_t}$ if the event occurs and $|\alpha_t-\beta_t|\cdot\frac{\alpha_t}{1-\alpha_t}$ if the event does not occur. %\lthought{I think it might be useful to define the payout in the intro, before we do the examples}.
%The payout for a incorrect bet is $-|\alpha_t-\beta_t|$.  

Suppose that $p_t$ is the true probability that an event occurs.  If player B bets that the event will occur, the expected payout is 
\[p_t \cdot|\alpha_t-\beta_t|\cdot (1-\alpha_t) - (1-p_t)\cdot|\alpha_t-\beta_t|\alpha_t = |\alpha_t-\beta_t|\cdot(p_t-\alpha_t).\]  This value is positive if $p_t >\alpha_t$;  that is, when the true probability of the event occurring is higher than the probability estimated by model $\alpha$.  If instead player B bets that the event will not occur, the expected payout becomes 
\[-p_t \cdot|\alpha_t-\beta_t|\cdot (1-\alpha_t) + (1-p_t)\cdot|\alpha_t-\beta_t|\alpha_t = |\alpha_t-\beta_t|\cdot(\alpha_t-p_t).\]  This value is positive if $p_t <\alpha_t$; that is, the true probability of the event occurring is lower than model $\alpha$'s estimate.  If player B correctly identifies whether model $\alpha$ is overestimating or underestimating the true probability, then the expected payout is positive. In either case, the magnitude of the expected payout is proportional to the distance between model $\alpha$'s estimate and the true probability.

%We chose these values as our payouts for two reasons: first, if $\alpha$ is perfectly calibrated, the expected payout from a single bet is $0$, second, the magnitude of payout increase as $|\alpha_t-\beta_t|$ increases.  

{
\renewcommand{\arraystretch}{1.3}
\begin{table}[htbp]
    \centering
    %\begin{tabular}{| r | c  c |}
    %\begin{tabular*}{0.5\linewidth}{@{\extracolsep{\fill}}|r|cc|}
    %    \hline
    %      & Event Occurs & Event Does Not Occur \\ \hline
    %     $\beta_t > \alpha_t$ & $|\alpha_t-\beta_t|\cdot(1-\alpha_t)$ &  $- |\alpha_t-\beta_t|\cdot(1-\alpha_t)$ \\ 
    %     $\beta_t \leq \alpha_t $ & $- |\alpha_t-\beta_t|\cdot\alpha_t$ & $|\alpha_t-\beta_t|\cdot\alpha_t$ \\
    %     \hline
%
%    \end{tabular*}  
        \begin{tabular*}{0.5\linewidth}{@{\extracolsep{\fill}}|r|cc|}
        \hline
          & Event Occurs & Event Does Not Occur \\ \hline
         $\beta_t > \alpha_t$ & $|\alpha_t-\beta_t|\cdot(1-\alpha_t)$ &  $- |\alpha_t-\beta_t|\cdot \alpha_t$ \\ 
         $\beta_t \leq \alpha_t $ & $- |\alpha_t-\beta_t|\cdot(1-\alpha_t)$ & $|\alpha_t-\beta_t|\cdot\alpha_t$ \\
         \hline

    \end{tabular*}  
    %\end{tabular}
    \caption{The payout in the betting game where model $\alpha$ is used to set the odds and model $\beta$ is used to determine the bet.}
    \label{tab:payouts}
\end{table}
}

%\lthought{I am a bit lost on the last two sentences here. I also thought in this first paragraph we could try adding the payout definitions before we use them in the examples.}

In the betting game, a bet is made by player B at the start of each round.  If the net payout is positive after summing over all rounds, %\lthought{We could try to be specific here and state how many betting rounds we are planning to have for our models comparison}
then model $\beta$ is said to have a \emph{betting edge} over model $\alpha$’s odds. Because the payouts depend on which model provides the odds, it is possible for both models to have a betting edge over the other’s odds (or for neither to have one). In such cases, we consider the model that achieves the higher overall profit to be superior.

To make sense of betting games, let us consider an example where Alice plays Bob in a game of Darts 271 as seen in Table \ref{tab:example game}.
{
\renewcommand{\arraystretch}{1.2}
\begin{table}[htbp]
    \centering
    \begin{tabular}{ |cc || c | c |}
        \hline
         Alice & Bob & $\alpha$ & $\beta$ \\
         \hline 
          0& 0 & 0.55 &  0.60 \\
          
          100& 120 & 0.52 &  0.40 \\
          
          250 & 275 & 0 & 0 \\
          \hline
    \end{tabular}
    \caption{An example darts game between Alice and Bob. Models $\alpha$ and $\beta$ estimate the likelihood that Alice wins.}
    \label{tab:example game}
\end{table}
}
The game begins with model $\alpha$ estimating that Alice has a $55\%$ probability of defeating Bob and model $\beta$ estimates that she has a $60\%$ probability of winning.  Therefore, in the betting game, player B bets that Alice will win the game since the estimation from model $\beta$ is greater than the estimation from model $\alpha$.  
%If she does, the payout is $|.60-.55|\cdot\frac{.45}{.55} = \frac{9}{220}$; if it does not occur, the payout is $-|.60-.55|=-.05$.  
A round of darts is played and the models update their estimations.  In this example, model $\alpha$ now estimates that the probability of Alice winning the game is $52\%$.  Model $\beta$ gives an estimation of $40\%$.  Since the estimation from model $\beta$ is less than the estimation from model $\alpha$, player B now places a bet that Alice will lose the game.  A final round of darts is played and Alice does indeed lose to Bob.  This means that player B's first bet was incorrect and their second bet was correct.  The payout for the first bet is $-|0.55-0.60|\cdot0.55=-0.0275$.  The payout for the second bet is  $|0.52-0.40|\cdot 0.52 = 0.0624$.  This means that player B earns a profit of $-0.0275 + 0.0624 = 0.0349$ from making these two bets.  In this small example, model $\beta$ has a betting edge over model $\alpha$'s odds.  A similar calculation to above can show that player A would have a net payout of $0.03-0.048=-0.018$ %\lthought{So I am getting the .03, but not getting the -.048. Here is what I have: 1st round: $.05*0.6 = 0.3$ and for 2d I have: $-0.12*(1-0.4) = -0.072$. I could be using the wrong equations though, but for the first round I am using the 0-0 weight and 1-0 for the 2d round} 
if the roles of model $\alpha$ and $\beta$ were reversed.  This means that model $\beta$ is superior to model $\alpha$. %\textbf{WE NEED TO DOUBLE CHECK THESE VALUES!!}

\section{Models}\label{sec: Models}

In this section, we describe the models developed to estimate a player’s probability of winning the game based on the current score. Each model was trained using only data from games completed before April 1, 2025. Conceptually, each model can be viewed as a function that takes two player IDs and their respective scores as inputs and returns a number between 0 and 1, indicating the estimated probability that the first player wins the game.

\subsection{The Null Model}
The first model that we consider is a model that only depends on the current game score and not the players involved in the game.  The \emph{null model}, $\pi_\text{NULL}$, estimates the probability that player $p_1$ wins based solely on how large player $p_1$'s lead or deficit is. 
Let $s_1$ and $s_2$ denote the current scores of players $p_1$ and $p_2$, respectively. The model’s estimated probability that player $p_1$ defeats player $p_2$ is given by
\[\pi_\text{NULL}(p_1, p_2, s_1, s_2) = \frac{1}{2}\Bigl(1 + \frac{s_1 - s_2}{100}\Bigr),\]
truncated to the interval $[0, 1]$ so the probabilities remain valid.  Note that if $|s_1-s_2|\leq100$, then $s_1-s_2 = 100\cdot\big(\pi_{\text{NULL}}(p_1, p_2, s_1, s_2) - \pi_{\text{NULL}}(p_2,p_1, s_2, s_1)\big)$. In plain words, the difference in score directly translates to the difference in each player’s probability of winning, expressed as a percentage.

For example, if player 1 leads 75 to 45, then the probability that player 1 wins is 30 percentage points higher than the probability that player 2 wins. That is to say, player 1 has a $65\%$ chance of winning, while player 2’s chance is $35\%$. %\lthought{Could clarify here a bit for example as ``..., numbers that add up to 100\% with a difference of 30\%"} \wthought{I think it's clear without the clarification.  We can let the readers do a little mental math if they want to.}
The choice to divide the score difference by 100 %\st{has no formal mathematical or statistical justification; it} 
was selected simply to make the score difference translate conveniently into a probability difference.

\subsection{Logistic Regression Model}
In the previous model, each additional point in the score difference increases the estimated probability of player $p_1$ winning by 0.5 percentage points. However, this approach fails to account for the fact that the importance of a single point depends on the game context. In a close match or early in the game, a one-point difference can have a substantial impact on the outcome, whereas in a lopsided game nearing completion, say, with a score of 250 to 190, a single point is unlikely to change the eventual winner. To address this limitation, 
we used a \emph{logistic regression model}, $\pi_{\text{REG}}$. Logistic regression estimates the probability of an event occurring by applying a logistic function to a linear combination of predictor variables. 
In our case, the predictors include the player IDs for both competitors, $p_1$ and $p_2$, and their current scores, $s_1$ and $s_2$ at the start of the round. 
The estimated probability that player 1 wins is
\[\pi_{REG}(p_1, p_2, s_1, s_2) = \frac{1}{1+e^{-(c_0+c_{p_1} - c_{p_2} +c_{1} \cdot s_1 + c_2 \cdot s_2)}}.\]
Here, $c_{p_1}$and $c_{p_2}$
 are the coefficients associated with the two players,  while $c_1$ and $c_2$ measure how each player’s current score influences the probability of player $p_1$ ultimately winning. These coefficients are estimated from the training dataset consisting of all games completed before April 1, 2025. %\lthought{So what exactly are $c_0, c_1,$ etc? Are they just coefficients that stay the same no matter the players $p_1$ and $p_2$ involved?} \athought{$c0, c_{s1}, c_{s2}$ are fixed for all players, but $c_{p1}, c_{p2}$ are not.}

The logistic regression model learns relationships between player identities and score values, estimating how strongly each feature contributes to the likelihood of winning a game. However, this approach still has certain limitations. If the dataset is not sufficiently large or diverse, the model may become overfit, essentially memorizing patterns rather than generalizing them, and may also carry historical bias into future predictions.  For an overview of implementing logistic regression in Python, see \cite{uvaLogisticRegression}. 

%In some cases, it memorized patterns (overfitting) rather than dynamically reacting to the game’s progression. It lacks real-time adaptability and can carry historical bias into new games.

\subsection{Basic Simulation Model}

Our dataset contains the point value of every recorded throw made by each player over the course of the tournament. This allows us to determine each player’s \emph{empirical point distribution} for a single throw.  This distribution gives the likelihood that a randomly sampled throw from a player scores a specific value.

%\begin{figure}
%    \centering
%    \includegraphics[width=0.9\linewidth]{Player_Throw_Distribution_SimilarMeans.png}
%    \caption{Score distribution graph}
%    \label{fig:player_throw_distribution_similar_mean}
%\end{figure}

%\begin{figure}
%    \centering
%    \includegraphics[width=0.9\linewidth]{Throw_Distribution_15.png}
%    \caption{Score distribution graph}
%    \label{fig:player_throw_distribution_20}
%\end{figure}

%\begin{figure}
%    \centering
%    \includegraphics[width=0.9\linewidth]{Throw_Distribution_2.png}
%    \caption{Score distribution graph}
%    \label{fig:player_throw_distribution_19}
%\end{figure}

%\begin{figure}
%    \centering
%    \includegraphics[width=0.9\linewidth]{Throw_Distribution_19.png}
%    \caption{Score distribution graph}
%    \label{fig:player_throw_distribution_random}
%\end{figure}

\begin{figure}
    \centering
    \includegraphics[width=0.90\linewidth]{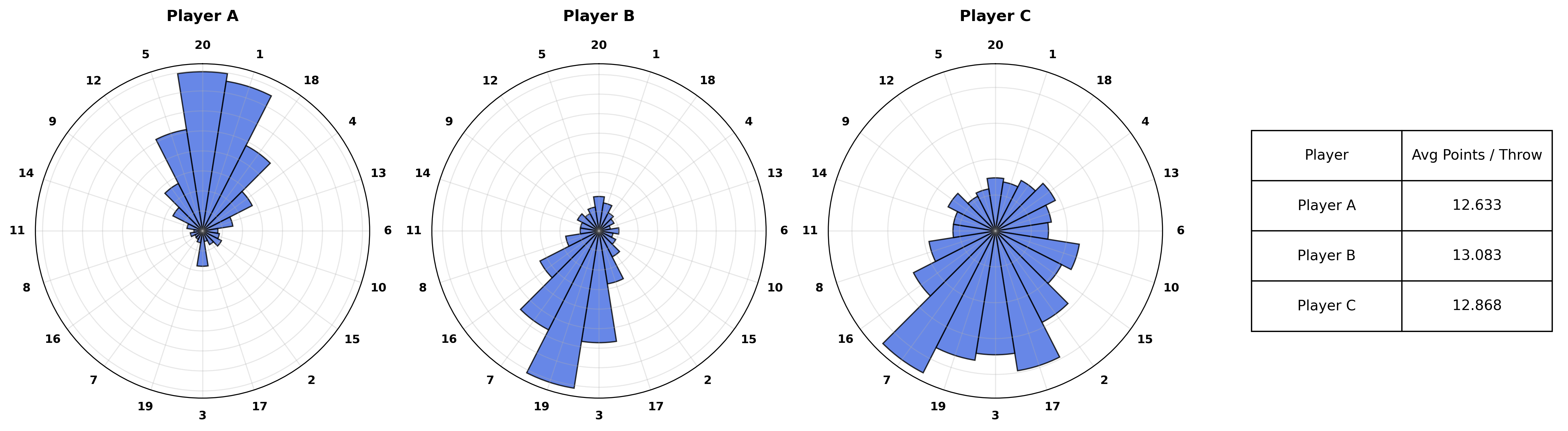}
    \caption{Throw distribution for sample players.}
    \label{fig:player_throw_distribution}
\end{figure}

Figure \ref{fig:player_throw_distribution} displays the throw distributions of three representative players. Based on these distributions, it appears that player A  typically targets the 20-point sector, player B favors the 19-point sector, and player C either varies their strategy or frequently aims for the bullseye.  These three players score a similar number of points per throw on average, but employ different strategies to achieve their average.  

It is theoretically possible that the relationships among these strategies are non-transitive, that is, for specific players, the strategy of aiming for the 20  on average outperforms aiming for the 19, aiming for the 19 strategy outperforms aiming for the bullseye, yet aiming for the 20 does not outperform the aiming for the bullseye approach.  See \cite{savage1994paradox} for an example of three dice that satisfy this non-transitive relation. A familiar example of a non-transitive game is Rock–Paper–Scissors, in which no single strategy dominates the others. The potential non-transitivity of Darts 271 implies that even when players’ point distributions are known, it can still be difficult to assess their relative strength directly and estimate the probability that one player defeats another.  

To address this challenge, we use Monte Carlo simulation to estimate the probability that one player defeats another player given the current score of the game. These simulations form the basis of our \emph{basic simulation model}, denoted by $\pi_{\text{SIM}}$.  Let $\pi_{\text{SIM}}(p_1, p_2, s_1, s_2)$ denote the probability that player $p_1$ ultimately wins in a game against player $p_2$ when the current score is $s_1$ to $s_2$.
To estimate this probability, we simulate $1000$ games beginning from the given score $s_1$ to $s_2$, using each player’s empirical point distribution that we derived from the dataset.  In each simulated game, players alternate turns, and on each turn a player receives three point values sampled independently from their empirical point distributions. This process repeats until both players have the same number of turns and one player is leading with a cumulative score that exceeds 271. At that point that player wins the simulated game.

If $W$ denotes the number of simulated games beginning with scores $s_1$ to $s_2$ won by player $p_1$, then the estimated probability that player $p_1$ defeats player $p_2$ is given by $\pi_{\text{SIM}}(p_1, p_2, s_1, s_2) = \frac{W}{1000}$.  That is to say, the estimated probability that player $p_1$ wins is the fraction of simulated games in which player $p_1$ is victorious.  We should note that this model faces some limitations as the given probability  is based on a random process which means it does not give a fixed probability value. However, by running 1000 simulations per scenario, these fluctuations are small enough to have minimal effect on the overall results.

\subsection{Adjusted Simulation Model}\label{Subsect Adjusted Sim}
Recall that the players in our dataset are amateur participants, many of whom had little to no prior experience throwing darts. As a result, it is likely that several, if not all, players had their skills change over the course of the tournament. The basic simulation model discussed in the previous section, however, weighs every throw equally, regardless of when in the season it occurred. As a result, if a player initially struggled to hit where they were aiming but later developed greater accuracy, the proportion of low-point throws in their empirical point distribution would overstate the current likelihood of missing their target.  Similarly, the player's empirical point distribution may underestimate their current likelihood of hitting the region they are aiming for.
%\st{The basic simulation model captured each player’s overall scoring distribution over the whole season, however, it fails when player's strategy shifts away from historical averages as players don't usually stick to one strategy especially in tournaments. Also, it is worth mentioning again that players are new to darts and so many of them appeared to improve a lot over the course of the semester.}

Common strategies used by the players in Darts 271 are to aim for the 20-point sector or the 19-point sector.  While some players aimed to just hit the board, this strategy was not commonly used by players that had confidence in their throws.  In fact, \cite{TibshiraniDarts} show that aiming for the triple 20 or the triple 19 regions are the best strategies for players that have some control over where the dart will land.  They found that aiming for the bullseye is best for players with low accuracy.  When building our \emph{adjusted simulation model}, we worked under the assumption that players aim for either the 20-point sector or the 19-point sector on each throw.  In this model, each player is characterized by two parameter vectors: one estimating their accuracy and another estimating the likelihood that their dart lands in a single, double, or triple region or off the board completely. Using these vectors, we simulate games in a similar manner as in the basic simulation model. The remainder of this section details how these two vectors are determined for each player and how the games are simulated.
%\st{In darts, a common strategy is to either aim for 20 or 19, but again, the choice of 20 vs 19 might change from throw to throw depending on the player's style. For that reason, we came up with an idea of combining both of these sections (19 \& 20) to be the target section which will be used to model our distance approach. }

\subsubsection{Player Distance Vector} \label{subsect: player distance vector}
\begin{figure}
    \centering
    \begin{tikzpicture}
  % radius of the circle
  \def\r{2.5cm}

  % sectors: index/color
  \fill[DistMiss!70,line width=0.6pt] (0,0) circle (\r);
  \foreach \i/\col in {0/DistMiss,1/DistMiss,2/Dist2,3/Dist1,4/Dist0,5/Dist1,
  6/Dist2,7/DistMiss,8/DistMiss,9/DistMiss,10/DistMiss,11/Dist2,
  12/Dist1,13/Dist0,14/Dist1,15/Dist2,16/DistMiss,17/DistMiss,
  18/DistMiss,19/DistMiss} {
    \pgfmathsetmacro\start{\i*18 + 9}
    \pgfmathsetmacro\End{(\i+1)*18 + 9}
    % fill sector
    \fill[\col!70] (0,0) -- (\start:\r) arc (\start:\End:\r) -- cycle;
  }

  \foreach \i in {2, 3, 4, 5, 6, 7, 11, 12, 13, 14, 15, 16}
    \pgfmathsetmacro\start{\i*18 + 9}
    \draw[black] (0,0) -- (\start:\r);
    \foreach \i in {0, 1, 8, 9, 10, 17, 18, 19}
    \pgfmathsetmacro\start{\i*18 + 9}
    %\draw[black!75, line width=0.05pt] (0,0) -- (\start:\r);
  % outer circle border
  \draw[black,line width=0.6pt] (0,0) circle (\r);
  \fill[DistMiss!70,line width=0.6pt] (0,0) circle (.094*\r);
  \draw[black,line width=0.6pt] (44:.094*\r) arc (44:136:.094*\r);
  \draw[black,line width=0.6pt] (206:.094*\r) arc (206:298:.094*\r);
  %\draw[black,line width=0.6pt] (0,0) circle (.95*\r);

  % optional: place labels in each sector
  \foreach \i/\label/\LABEL in {2/2/18,3/1/1,4/0/20,5/1/5, 
  6/2/12,11/2/16, 12/1/7,13/0/19,14/1/3,15/2/17, 19/MISS/} {
    \pgfmathsetmacro\ang{(\i*18 + (\i+1)*18)/2 + 9} % mid-angle
    \node at (\ang:1.6cm) {\small\textbf{\label}};
    \node at (\ang:\r+.25cm) {\footnotesize{\LABEL}};
  }
  \node at (171:1.6cm) {\small\textbf{MISS}};
\end{tikzpicture}
    \caption{Each sector is classified as Distance 0, Distance 1, Distance 2, or Miss.}
    \label{fig:Dist Values}
\end{figure}
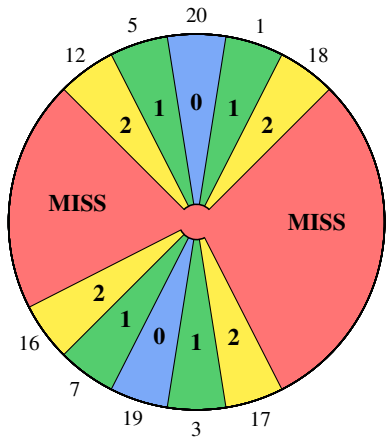

We developed a method to estimate a player’s accuracy as a function of time by first classifying each throw into one of four accuracy categories: \emph{Distance 0}, \emph{Distance 1}, \emph{Distance 2}, or \emph{Miss}. These categories represent how close the throw was to the intended target.  See Figure \ref{fig:dart_board} as a reminder of the arrangement of the point sectors of a dartboard. The following scoring values were assigned to each category:
\begin{itemize}
\item \textbf{Distance 0:} 19, 20, 38 ($2\times19$), 40 ($2\times20$), 57 ($3\times19$), 60 ($3\times20$); %\lthought{I think it would be nice to clarify here that 38 comes from 19*2, 40 - from 20*2, 57 - from 19*3, etc}
\item \textbf{Distance 1:} 1, 3, 5, 7, 21 ($3\times7$);
\item \textbf{Distance 2:} 12, 16, 17, 18, 24 ($2\times12$), 32 ($2\times16$), 34 ($2\times17$), 36 ($2\times18$), 48 ($3\times16$), 51 ($3\times7$), 54 ($3\times18$);
\item \textbf{Miss:} all other scores.
\end{itemize}
We note that our dataset does not distinguish between scores that can be achieved in multiple ways. For example, a score of 18 could result from hitting the single 18, double 9, or triple 6 regions. In our analysis, we treat each such score as originating from the single region. That is, every recorded score (e.g., 18) is assumed to represent a throw that hit the corresponding single region, not a double or triple of another number.  
For that reason, the scores $2$, $6$, $9$, $10$, $14$, and $15$ are not classified as Distance 1 despite the fact that those scores could theoretically occur by missing the intended target by one sector and landing in the double or triple region.  We chose to classify the score of 24 as Distance 2.  Since only $0.2\%$ of all throws in the dataset were worth 24 points, classifying it as Miss could be justified. %\lthought{I think we need to clarify how we know that it is a rare score.} \wthought{Can you look up how many 24 point throws occurred throughout the whole season?} \athought{I found 99 throws}  
Figure \ref{fig:Dist Values} provides a visual of the classification of each sector of the dartboard.

%\st{As a result, we came up with a new simulation model that accounts for:}
%\begin{enumerate}
%    \item \st{The distance of each throw from a target (e.g., the 19/20 region, where players commonly aim)}\cite{TibshiraniDarts}
%    \item \st{The probabilities of hitting multipliers (single, double, or triple rings).}
%\end{enumerate}
%\st{By incorporating these, the model simulates gameplay in a way that mirrors human decision-making and skill variation (We capture player style and strategy over time).}

We constructed a dataset in which each row corresponds to a single player in a specific game. For each of those games, we recorded the player ID, the date and time of the game, and the observed proportions of throws classified as Distance 0, Distance 1, Distance 2, and Miss. For each player, we then fit four separate linear regression models, one for each accuracy category, using date and time as the predictor variable. These regression models were used to estimate each player’s expected proportion of throws classified as Distance 0, Distance 1, Distance 2, and Miss on April 1, 2025. Because the resulting estimates do not necessarily sum up to 1, we normalize each player’s estimated proportion vector by dividing the vector by the total sum of its values.  The resulting vector is called the player's \emph{distance vector}. %\textbf{SHOULD IT BE VECTOR OR DISTRIBUTION?}.

\begin{figure}[htbp]
    \centering
    \includegraphics[width=0.9\linewidth]{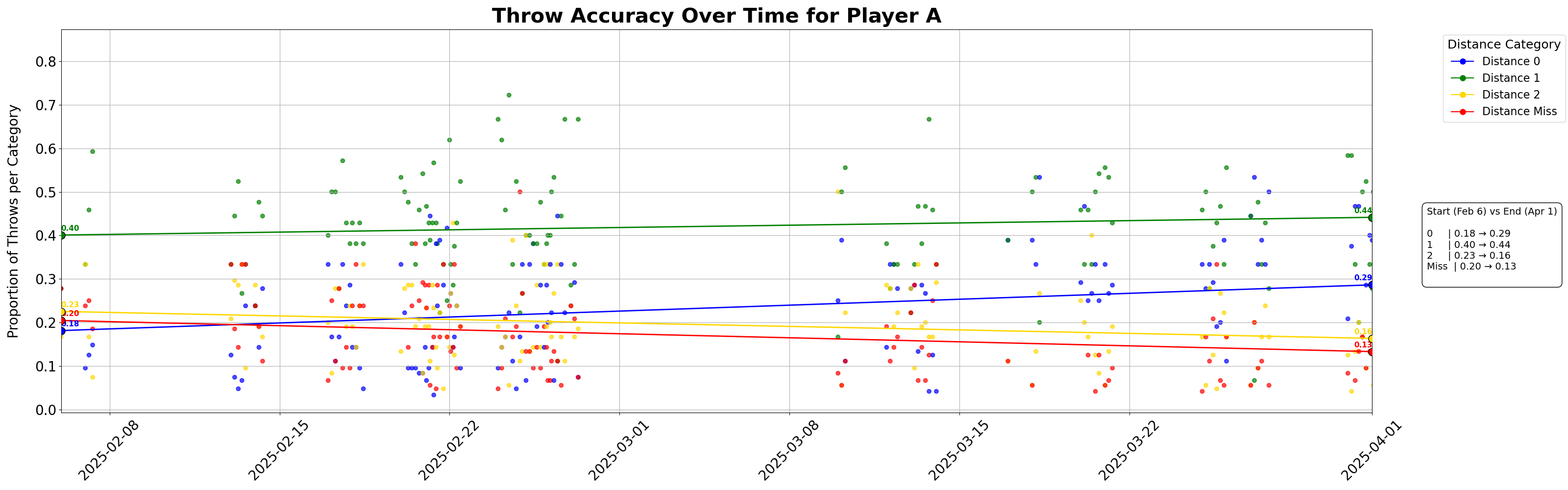}
    \caption{Change in distribution vectors throughout the season.}
    \label{fig:distance_from_target}
\end{figure}

For clarification, consider the scatterplot shown in Figure \ref{fig:distance_from_target}. 
%\lthought{I think we should consider increasing the font of the titles and the axes labels and, if possible, making the style of this graph align with the other ones in Figure 6.}  
Each game played is represented by four dots on the scatterplot above the start date and time of the game.  For a specific game, the height of the blue dot associated with it is the proportion of Distance 0 throws to the total number of throws, the height of the green dot is the proportion of Distance 1 throws, the height of the yellow dot is the proportion of Distance 2 throws, and the height of the red dot is the proportion of throws classified as Miss.  For the player shown in Figure \ref{fig:distance_from_target}, we see that the proportion of Distance 0 throws appears to be increasing over time, meaning that the player is improving their accuracy.  This player has a distance vector of \[\frac{1}{0.29+0.44+0.16+0.13}\langle 0.29, 0.44, 0.16, 0.13 \rangle =\langle 0.284, 0.431, 0.159, 0.127\rangle.\] 
We interpret these values in the following way: if a game was played after April 1, 2025, then we would expect $28.4\%$ of this player's throws to land in the 20-point or 19-point sector, $43.1\%$ of their throws to land one sector away from their target, $15.9\%$ of their throws to land two sectors away from their target, and the remaining $12.7\%$ of their throws will land more than two sectors away from their target.

When simulating games for the adjusted simulation model, we used the the player's distance vector as opposed to their empirical point distribution to determine the number of points the player earned on a specific throw.  We discuss this process in detail in Section~\ref{subsub: Sim Games}

\subsubsection{Player Multiplier Vector}
When a player aims, it is in their best interest to target the region of the sector that triples the point value.  However, as discussed above, players do not always hit their intended target.  For any given throw, the possible multipliers are $0\times$, $1\times$, $2\times$, and $3\times$.  A multiplier of $0\times$ indicates that the dart did not land on the board.  Because our dataset does not record which multiplier was applied on each throw, we instead estimate these values.  To do so, for each player we define a \emph{multiplier vector} 
\[\mathbf{m} = \langle m_0, m_1, m_2, m_3\rangle,\]
where each $m_k$ represents the estimated proportion of throws that received a multiplier of $k\times$.  The value of $m_0$ is taken as the empirical proportion of the player's throws that scored zero points.  The remaining probabilities are scaled to sum to $1-m_0$, representing the throws that landed on the board.  

Let $T$ be a set and let $N_T$ denote the number of player's throws whose value is in the set $T$.  Let $S$ be the set of values that are classified as Distance 0.  That is to say $S = \{19, 20, 38, 40, 57, 60\}$.  Then $N_S$ is the total number of a player's throws that are classified as Distance $0$.  We define the values of $m_1$, $m_2$, and $m_3$ by, 
\[m_1 = (1-m_0)\cdot \frac{N_{\{19, 20\}}}{N_S}, \hspace{1 cm} m_2 = (1-m_0)\cdot \frac{N_{\{38, 40\}}}{N_S}, \hspace{1 cm} m_3 = (1-m_0)\cdot \frac{N_{\{57, 60\}}}{N_S}.\]
This means that $m_1$, $m_2$, and $m_3$ are relative frequencies of hitting the single, double, and triple regions among all darts that landed within the 19- and 20-point sectors.  It is worth pointing out that we make a simplifying assumption that the proportion of doubles and triples observed in these sectors accurately reflects the player's overall likelihood of hitting double or triple regions elsewhere on the board. %\lthought{I'm confused on how this last sentence is relevant.} \wthought{It is an underlying assumption used in the model.}

\subsubsection{Simulating Games}\label{subsub: Sim Games}
As with the basic simulation model, the foundation of the adjusted simulation model is a Monte Carlo simulation. In this model, each simulated throw consists of two components: a point value determined by the player’s distance vector and a multiplier value determined by the player’s multiplier vector. Before applying the multiplier, we assign baseline point values to each distance category as follows: 19.5 for Distance 0 throws, 4 for Distance 1 throws, 15.75 for Distance 2 throws, and 9.2 for Miss throws. These baseline values were computed by averaging the scores of the sectors corresponding to each distance category. %\lthought{Averaging out across the whole dataset?} \wthought{No.  The two Distance 0 sectors are 20 and 19 so Distance 0 have the value of 19.5.  The four Distance 1 sectors are 1, 3, 5, 7 so Distance 1 has the value of 4 etc.}

Let $\pi_{\text{ADJ}}(p_1, p_2, s_1, s_2)$ denote the probability that player $p_1$ ultimately wins in a game against player $p_2$ when the current score is $s_1$ to $s_2$.
Like the basic simulation model, to estimate this probability, we simulate $1000$ games beginning from the given score $s_1$ to $s_2$.  Instead of using each player's empirical point distribution, we use their distance vector and multiplier vector to determine the point value of a throw.  In each simulated game, players alternate turns, and on each turn a player receives three point values, each determined as the product of two random components: a baseline point value sampled from the player’s distance vector and a multiplier sampled from the player’s multiplier vector. %\lthought{This is a repetition from the Basic Simulation section.}

Formally, if $v_t$ is the baseline value drawn from the distance vector and $m_t$ is the multiplier drawn from the multiplier vector for throw $t$, then the score for that throw is $s_t = v_t\cdot m_t$.  Like with the basic simulation mode, this process repeats until both players have the same number of turns and one player is leading with a cumulative score that exceeds 271. At that point that player is declared the winner of the simulated game. %\lthought{I think we should not be repeating the same sentences from the Basic Simulation model section. We could try referencing it and explaining that it works the same way or rewording this paragraph.}

The estimated probability that player $p_1$ ultimately wins is determined exactly as it is with the basic simulation model.  That is to say, if $W$ denotes the number of simulated games beginning with the score $s_1$ to $s_2$ won by player $p_1$, then the estimated probability that player $p_1$ defeats player $p_2$ is given by $\pi_{\text{ADJ}}(p_1, p_2, s_1, s_2) = \frac{W}{1000}$. We should note that this model faces the same limitations as the basic simulation model as the probability given is based on a random process which means it does not give a fixed probability value. %However, as with the basic simulation model, by running 1000 simulations per scenario, these fluctuations are small enough to have minimal effect on the the overall results. %\lthought{We should not be copying the same paragraph from Basic Simulation section into here.}

\subsection{Score-Dependent Massey Model} \label{SubSect: SDMM}

Our final model is a variation of a well-known linear algebra based model.  In the 1990s, \cite{minton_1992} and \cite{Massey1997} independently introduced the idea of using a system of linear equations to rate sports teams. In this method, which is commonly referred to as the \emph{Massey method}, each game played corresponds to an equation in a system of equations. If team $i$ defeats team $j$, the equation $r_i - r_j = 1$ is included in the system, while a loss for team $i$ corresponds to $r_i - r_j = -1$.  The system can be represented in matrix form as $X\mathbf{r}=\mathbf{y}$, where $\mathbf{r}$ contains the team ratings. The least-squares solution to this system yields the ratings of the teams. After many games have been played, the difference $r_i-r_j$ can be interpreted as an estimate of the difference in the winning probabilities of team $i$ and team $j$
when they face each other.

The model described above can only be used to predict the outcome at the start of a game. However, our models must be able to make predictions after play has begun. To address this, we propose a new approach that incorporates the current score. As in the traditional Massey method, we construct a system of linear equations, but instead of having one equation per game, we include one equation for each round. In addition, each player is assigned two parameters, which we will refer to as ratings, that are combined with weights to estimate their \emph{perceived strength} at any point in the game. We denote as $r_i^{(1)}$ and $r_i^{(2)}$ the two ratings associated with player $p_i$ and assign the value of $\frac{271-s_i}{271}r_i^{(1)} + \frac{s_i}{271}r_i^{(2)}$ to be the perceived strength of player $p_i$ when their current score is $s_i$.

If player $p_i$ led player $p_j$ by a score of $s_i$ to $s_j$ at the start of a round and the game ended with player $p_i$ being victorious, then the equation
\[
\frac{271-s_i}{271}r_i^{(1)} + \frac{s_i}{271}r_i^{(2)} 
- \left( \frac{271-s_j}{271}r_j^{(1)} + \frac{s_j}{271}r_j^{(2)} \right) = 1
\]
is included in the system of equations. If player $p_i$ instead lost the game, the corresponding equation
\[
\frac{271-s_i}{271}r_i^{(1)} + \frac{s_i}{271}r_i^{(2)} 
- \left( \frac{271-s_j}{271}r_j^{(1)} + \frac{s_j}{271}r_j^{(2)} \right) = -1
\]
is included. In the rare situation where a game is tied and both players have a score above 271, we use the value of 271 instead of the current score. These equations can be interpreted as saying that the difference in perceived strengths at the start of the round is equal to $1$ or $-1$ depending on which player ultimately wins the game. Once the game concludes, we add the equation
\[
r_i^{(2)} - r_j^{(2)} = 1
\]
if player $p_i$ wins, or
\[
r_i^{(2)} - r_j^{(2)} = -1
\]
%\lthought{I think these equations are not quite right. If a game ended with, let's say 281-241, the equation that gets added to the matrix is $r_i^{(2)} - \frac{241}{271}r_j^{(2)} = \pm1$. So, if the score is above 271, we are just replacing it with 271 and then applying the regular weights to them.}
if player $p_i$ loses.  Therefore, each game of darts contributes one equation for each round in the game and an additional equation representing the conclusion of the game.

As an example, the game between Alice ($p_A$) and Bob ($p_B$) shown in Table~\ref{tab:example game} would contribute the following three equations to the system:

\[
\begin{array}{r c r c r c r c l}
r_A^{(1)} & & & - & r_B^{(1)} & &  &= & -1,  \\
\frac{171}{271}r_A^{(1)} & + & \frac{100}{271}r_A^{(2)} & - & \frac{151}{271}r_B^{(1)} & - & \frac{120}{271}r_B^{(2)} & = & -1,  \\
 & & r_A^{(2)} & && - & r_B^{(2)} & = & -1.
\end{array}
\]
The first two equations correspond to the two rounds played in the game, and the final equation represents the conclusion of the game. The $-1$ on the right-hand side of each equation indicates that Bob was victorious.

As with the classical Massey method, the values of each $r_i^{(1)}$ and $r_i^{(2)}$ are determined by finding the least-squares solution to the matrix equation $X\mathbf{r} = \mathbf{y}$. The matrix $X$ and vector $\mathbf{y}$ contain a row for each round and each game played. We also include additional rows that guarantee some desired properties.  If $n$ is the number of players in the field, we include an additional $6\cdot\binom{n}{2}$ rows in the matrix as follows. For $1 \leq i<j \leq n$, we include rows corresponding to the following linear equations:
\[
\begin{array}{r c r c r c r c r}
r_i^{(1)} & & & - & r_j^{(1)} & & &= & 1,  \\
\frac{1}{2}r_i^{(1)} & + & \frac{1}{2}r_i^{(2)} & - & \frac{1}{2}r_j^{(1)} & - & \frac{1}{2}r_j^{(2)} & = & 1,  \\
 & & r_i^{(2)} & & & - & r_j^{(2)} & = & 1, \\
 r_i^{(1)} & & & - & r_j^{(1)} & & &= & -1,  \\
\frac{1}{2}r_i^{(1)} & + & \frac{1}{2}r_i^{(2)} & - & \frac{1}{2}r_j^{(1)} & - & \frac{1}{2}r_j^{(2)} & = & -1,  \\
 & & r_i^{(2)} & & & - & r_j^{(2)} & = & -1.
\end{array}
\]
By including these extra equations, we guarantee that the matrix $X^\top X$ has nullity~1.  To ensure that the ratings are well-defined, we require that the sum of all ratings is $0$. This rating system has the additional property that players who have participated in only a few games will tend to have ratings close to the league average, since the supplementary linear equations effectively assume that each player both defeats and loses to every other player in closely played games. The \cite{colley_2002} method shares a similar property, where a player is deemed average until sufficient evidence is presented to show otherwise.

From here we define the \emph{score-dependent Massey model} (SDMM), $\pi_\text{SDMM}$ by 
\[\pi_\text{SDMM}(p_1, p_2, s_1, s_2) =\frac{1}{2}\Biggl(1+ \frac{271-s_1}{271} r_1^{(1)} + \frac{s_1}{271}r_1^{(2)} - \Bigl(\frac{271-s_2}{271}r_2^{(1)} + \frac{s_2}{271}r_2^{(2)}\Bigr)\Biggr), \]
truncated to the interval $[0, 1]$ so the probabilities remain valid.

As in the original Massey method, the difference in the estimated win probabilities between the two players can be interpreted as the difference in their perceived strengths at the start of the round. In particular,
\[
 \pi_\text{SDMM}(p_1, p_2, s_1, s_2) - \pi_\text{SDMM}(p_2, p_1, s_2, s_1) = \frac{271-s_1}{271}r_1^{(1)} + \frac{s_1}{271}r_1^{(2)} 
- \left( \frac{271-s_2}{271}r_2^{(1)} + \frac{s_2}{271}r_2^{(2)} \right).
\]

Note that, as a consequence of this formulation, we can directly estimate the marginal effect of scoring one additional point on an average player’s probability of winning. Let $r^{(1)}$ and $r^{(2)}$ be the average value of $\Bigl\{r_i^{(1)}\Bigr\}$ and $\Bigl\{r_i^{(2)}\Bigr\}$ respectively and define $\Delta r = r^{(2)}- r^{(1)}$.   Then, if player $p_1$ and player $p_2$ both have the league average values for their two ratings, we have 
\begin{align*}
\pi_\text{SDMM}(p_1, p_2, s_1, s_2) &= \frac{1}{2}\Biggl(1+\frac{271-s_1}{271}r^{(1)} + \frac{s_1}{271}r^{(2)} 
- \Bigl( \frac{271-s_2}{271}r^{(1)} + \frac{s_2}{271}r^{(2)} \Bigr)\Biggr) \\
 &= \frac{1}{2}+\frac{s_2-s_1}{542}r^{(1)} + \frac{s_1-s_2}{542}(r^{(1)}+\Delta r) \\
 & = \frac{1}{2} + \frac{\Delta r}{542}(s_1-s_2).
\end{align*}
That says that if two league-average players compete against each other, then each additional point scored increases the player’s estimated probability of winning by $\frac{100\cdot\Delta r}{542}$ percentage points.

%\lthought{To clarify how this model works, let us consider an example where Alice has $r_1=-1.3$ and $r_2=1.7$ and Bob has $r_1=-1.4$ and $r_2=1.8$. So, at the start of the game, when $r_1$ parameters carry more weight, the SDMM model predicts that Alice will do better than Bob since her $r_1$ is higher. But, as the game progresses, Bob is projected to do better since his $r_2$ is higher. Looking back at the game between Alice and Bob displayed in~\ref{tab:example game}, at the very beginning of the game, when the score is $0-0$, }

Let us consider an example to help clarify how this model works.  Suppose Alice has ratings of $r_A^{(1)} = -1.3$ and $r_A^{(2)} = 1.7$ and suppose Bob has ratings of $r_B^{(1)} = -1.4$ and $r_B^{(2)} = 1.8$. Since the first rating associated with Alice is greater than that of Bob's, the score-dependent Massey model will initially estimate Alice has a probability of winning that is $.1$ greater than Bob's.  This is because  $r_A^{(1)} - r_B^{(1)} = .1$.  However, since Alice's second rating is less than Bob's, if the game remains close, the model will eventually estimate that Bob is the more likely winner.  Using the scores from the game between Alice and Bob shown in Table~\ref{tab:example game}, we can see how the score-dependent Massey model adjusts its estimates as the game progresses.  At the start of the game, the estimated probability that Alice wins is
\[\pi_{\mathrm{SDMM}}(\text{Alice},\text{Bob},0,0) = \frac{1}{2}\Bigl(1-1.3-(-1.4)\Bigr) = 0.55.\]
At the start of the second round, the score is 100 points for Alice and 120 points for Bob. The model now estimates Alice’s probability of winning as
\[\pi_{\mathrm{SDMM}}(\text{Alice},\text{Bob},100,120) = \frac{1}{2}\Bigl(1+\frac{171}{271}\cdot(-1.3)+
\frac{100}{271}\cdot(1.7)-\Bigl(\frac{151}{271}\cdot(-1.4)+
\frac{120}{271}\cdot(1.8)\Bigr)\Bigr) = 0.395.\]
Even though the score remains relatively close, the model shifts to favor Bob, reflecting his higher second rating $r_B^{(2)}$, which becomes more influential as the game nears completion.

\section{Results}\label{sec: Results}
In this section we compare the performance of the five models under consideration in terms of Brier scores and head-to-head matchups in the betting game.  Recall that each model is trained using the games played before April 1, 2025, and evaluated using the games played on or after April 1, 2025. In addition, model performance is assessed based on estimations made at the start of each round, not just at the start of the game.

\subsection{Brier Score Results}\label{Subscect: Brier Score Reults}

Table \ref{tab:brier-six-col} shows the Brier score for each model when evaluated on all rounds played on or after April 1, 2025. Additionally, the table shows the Brier score for each model when evaluated at a fixed round number. Here, $n$ denotes the sample size. For example, there were $353$ games that reached a fifth round; the Brier scores in that row were computed using only the predictions made at the start of Round~5.  

\begingroup
\setlength{\tabcolsep}{12pt}
\renewcommand{\arraystretch}{1.35} 
\begin{table}[htbp]
\centering
\small
\begin{tabular*}{0.8\linewidth}{@{\extracolsep{\fill}}|r|rrrrr|}
\hline
 & Null  & Logistic & Basic Sim. & Adjusted Sim. & SDMM \\
\hline
Round 1 ($n = 362$)   & 0.2500 & 0.2571 & 0.2425 & 0.2443 & \textbf{0.2372} \\
Round 2 ($n = 362$)   & 0.2247 & 0.2318 & 0.2208 & 0.2237 & \textbf{0.2135} \\
Round 3 ($n = 362$)   & 0.1992 & 0.2046 & 0.1983 & 0.1981 & \textbf{0.1899} \\
Round 4 ($n = 361$)   & 0.1759 & 0.1807 & 0.1729 & 0.1735 & \textbf{0.1678} \\
Round 5 ($n = 353$)   & 0.1498 & 0.1542 & 0.1445 & 0.1457 & \textbf{0.1419} \\
Round 6 ($n = 307$)   & 0.1502 & 0.1561 & 0.1426 & 0.1421 & \textbf{0.1418} \\
Round 7 ($n = 203$)   & 0.1666 & 0.1867 & 0.1541 & \textbf{0.1509} & 0.1573 \\
Round 8 ($n = 105$)   & 0.1937 & 0.2057 & 0.2013 & 0.2057 & \textbf{0.1834} \\
Round 9 ($n = 37$)    & 0.1613 & 0.1754 & \textbf{0.1232} & 0.1279 & 0.1545 \\
Round 10 ($n = 10$)   & 0.1902 & 0.2637 & 0.1786 & \textbf{0.1624} & 0.1910 \\
Round 11 ($n = 3$)    & 0.1577 & 0.2328 & 0.0983 & \textbf{0.0835} & 0.1463 \\
\hline
All rounds ($n = 2465$)   & 0.1902 & 0.1980 & 0.1849 & 0.1857 & \textbf{0.1807}  \\
\hline
\end{tabular*}
\caption{The Brier score for each model grouped by round number.}
\label{tab:brier-six-col}
\end{table}
\endgroup
%\lthought{Need to define what $n$ is in Table \ref{tab:brier-six-col}}

As shown in the table, the score-dependent Massey model performs best overall in terms of the Brier score. This result is not surprising, since the rating values used in the formula for $\pi_{\text{SDMM}}$ were obtained by solving a least-squares problem that minimizes the sum, over all rounds in the training set, of the squared differences between the predicted win-probabilities difference and the actual game outcomes.

Let $t$ denote a round in the training set, and let players $p_{t_1}$ and $p_{t_2}$ be the two players competing in round $t$. Let $s_{t_1}$ and $s_{t_2}$ denote their respective scores, and let $\sigma_t$ denote the game outcome (equal to 1 if $p_{t_1}$ wins and 0 otherwise). Then the least-squares problem determines rating values that minimize
\[\sum_t \Bigl(\pi_\text{SDMM}(p_{t_1}, p_{t_2}, s_{t_1}, s_{t_2}) - \pi_\text{SDMM}(p_{t_2}, p_{t_1}, s_{t_2}, s_{t_1}) - (\sigma_{t} - (1-\sigma_t)\Bigr)^2,\]
where the sum runs over all rounds in the training set.
Since $\pi_\text{SDMM}(p_{t_2}, p_{t_1}, s_{t_2}, s_{t_1}) = 1 - \pi_\text{SDMM}(p_{t_1}, p_{t_2}, s_{t_1}, s_{t_2})$, we can simplify this expression as
\[\sum_t \Bigl(\pi_\text{SDMM}(p_{t_1}, p_{t_2}, s_{t_1}, s_{t_2}) - \pi_\text{SDMM}(p_{t_2}, p_{t_1}, s_{t_2}, s_{t_1}) - (\sigma_{t} - (1-\sigma_t))\Bigr)^2 = 4\sum_t(\pi_\text{SDMM}(p_{t_1}, p_{t_2}, s_{t_1}, s_{t_2}) - \sigma_{t})^2.\]
This is exactly four times the Brier score of $\pi_{\text{SDMM}}$ when evaluated on the training set.  In other words, solving the least-squares problem directly minimizes the Brier score.

\begin{figure}[htbp]
    \centering
    \includegraphics[width=0.9\linewidth]{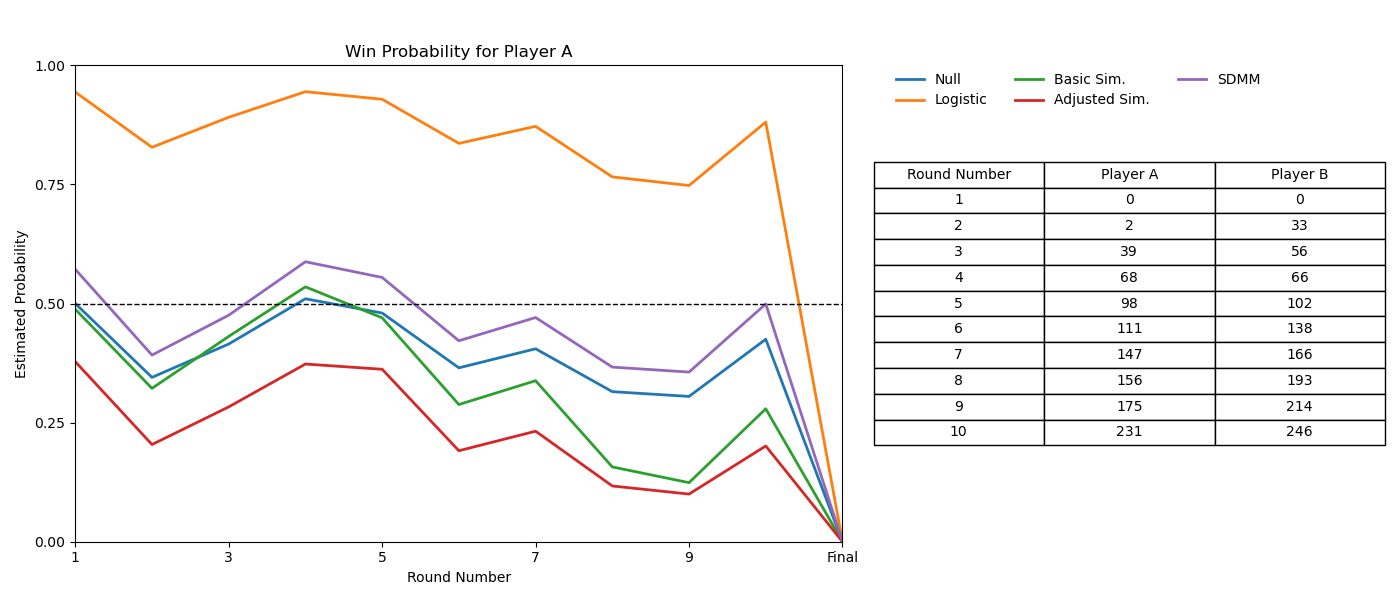}
    \caption{Predicted win probability by round for all five models in a game between two players.}
    \label{fig:Logistic Fail}
\end{figure}

Another observation from Table \ref{tab:brier-six-col} is that $\pi_{\text{REG}}$ performs worse than the null model in terms of Brier score. We suspect that the logistic regression model places too much weight on the outcomes of specific matchups in the training set and fails to account for changes that occur throughout the season.  That is, a game played early in the season may have limited predictive value for games played much later. This limitation likely arises because players improve (or decline) at different rates over the course of the season, causing the relationships learned by the regression model to become outdated.  Consider the game shown in Figure \ref{fig:Logistic Fail}. 
%\lthought{Despite trailing for the entire match, $\pi_{\text{REG}}$ consistently estimates that a student player, $p_S$, is favored to win. Toward the end of the game, the other four models correctly adjust to reflect that player Truong Le, $p_T$, is now expected to win. In the training set, these two players faced each other three times, and player $p_S$ won every matchup.}
Despite trailing for the entire match and scoring only two points in the first round, $\pi_{\text{REG}}$ consistently estimates that player A is heavily favored to win. Towards the end of the game, the other four models correctly adjust to reflect that player A is unlikely to win. This may be due to the fact that in the training set, player B played only eight games and won just one. 
As a result, the logistic regression model became overfit to this specific player, effectively learning that player B usually loses, regardless of the current game context.  Situations like the one shown in Figure \ref{fig:Logistic Fail} may help explain why the logistic model performs the worst in terms of Brier score.

As we just mentioned, the null model outperforms the logistic model in terms of Brier score.  We suspect that this happened by chance.  Recall that the estimated probability of player $p_1$ winning given by the null model when the score is $s_1$ to $s_2$ is given by \[\pi_\text{NULL}(p_1, p_2, s_1, s_2) = \frac{1}{2}\Bigl(1 + \frac{s_1 - s_2}{100}\Bigr).\]  This means that for each additional point scored by player $p_1$, the model estimates that the probability that player $p_1$ wins increases by $0.5$ percentage points.  We saw in Section \ref{SubSect: SDMM} that each additional point scored increases the player's estimated probability of winning by $\frac{100\Delta r}{542}$ percentage points, where $\Delta r = 3.176$ (see Table \ref{tab:SDMM Summary}) is the value obtained by solving a least-squares problem during model fitting for $\pi_{\text{SDMM}}$.  This means that the score-dependent Massey model estimates a marginal effect of $\frac{100\cdot 3.176}{542} = 0.59$ percentage points per point scored.  The fact that $0.5$ happens to be close to this value may explain why the null model performs comparatively well in terms of Brier score.  It turns out that if we were to define the null model as \[\pi'_\text{NULL}(p_1, p_2, s_1, s_2) = \frac{1}{2}\Bigl(1 + \frac{s_1 - s_2}{85}\Bigr),\] 
then each additional point scored increases the player's estimated win probability by $\frac{100}{85\cdot 2} = 0.59$ percentage points, the same marginal effect implied by the score-dependent Massey model. The Brier score for this model when evaluated over all rounds in the test set is $B(\pi'_{\text{NULL}}) = 0.1889$, a slight improvement over the Brier score of the null model.  %\lthought{which is comparable to the score of $0.1807$ calculated for the SDMM model. (I just felt like I wanted to clarify where 85 came from sooner).} 

%\st{This model satisfies that each additional point scored increases the player's estimated win probability by $\frac{100}{170} = 0.59$ percentage points, the same marginal effect implied by the score-dependent Massey model.}

\begin{table}[htbp]
\renewcommand{\arraystretch}{1.25}
    \centering
    \begin{tabular}{| r | c c c c c c c |}
        \hline
         Rating & Mean & SD & Median & Q1 & Q3 & Min & Max \\
          \hline
          $r^{(1)}$ & -1.588 & 0.088 & -1.603 & -1.634 & -1.548 & -1.738 & -1.323 \\
          $r^{(2)}$ & 1.588 & 0.072 & 1.577 & 1.555 & 1.635 & 1.421 & 1.784 \\
          \hline
    \end{tabular}
    \caption{Summary statistics of the two ratings used in the SDMM model.}
    \label{tab:SDMM Summary}
\end{table}
%\begingroup
%\setlength{\tabcolsep}{12pt}
%\renewcommand{\arraystretch}{1.35} 
%\begin{table}[htbp]
%\centering
%\small
%\begin{tabular}{lrr}
%\hline
% & Brier (All rounds) & Brier (Start-of-game) \\
%\hline
%Null Model     & 0.0000 & 0.0000 \\
%Basic Sim.     & 0.0000 & 0.0000 \\
%Adjusted Sim.  & 0.0000 & 0.0000 \\
%Logistic       & 0.0000 & 0.0000 \\
%SDMM & 0.1807 & 0.2398 \\
%\hline
%\end{tabular}
%\caption{Brier score table}
%\label{tab:brier-two-col}
%\end{table}
%\endgroup

\subsection{Betting Game Results}\label{Subsect Betting Game Results}
\begin{figure}[htbp]
    \centering
    \includegraphics[width=0.7\linewidth]{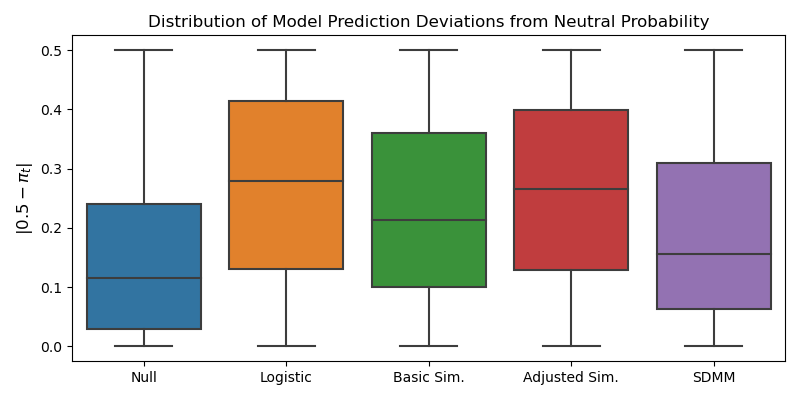}
    \caption{Distribution of prediction deviations from 0.5.}
    \label{fig:Distribution_of_model_prediction}
\end{figure}

Table \ref{tab:betting-all} and Table \ref{subtable Betting Game} show the results of the betting game when bets are placed at the start of each round. In this version of the game, the score-dependent Massey model outperforms all other models when compared head-to head. The adjusted simulation model earned the most profit when the logistic model was used to set the odds.  Figure \ref{fig:Distribution_of_model_prediction} shows the distribution of predictions made by each of the five models. The vertical axis represents the deviation of the estimated win probability from $50\%$. We see that the adjusted simulation model and the logistic regression model generate the most extreme predictions. We suspect that the reason why the adjusted simulation model has a superior performance compared to the logistic regression model is due to the overfitting issue discussed in Section \ref{Subscect: Brier Score Reults}.

One surprising result is that the basic simulation model outperforms the adjusted simulation model by 1.92 points in the head-to-head comparison. One possible explanation is that the methods described in Section
\ref{Subsect Adjusted Sim} do not properly capture the players’ improvement.    

%Note that most of the entries in Table \ref{tab:betting-all} are negative. This is a consequence of how payouts are assigned. Toward the end of a game, it is often the case that one player becomes a heavy favorite meaning the actual probability that player $p_1$ wins is close to $0$ or $1$. Making a bet under those conditions has limited upside and substantial downside. In the betting game, the model that gives the lower estimated probability for the favorite is required to predict that the favorite will lose. In the rare cases when the favorite does lose, the payout is large, but most often the favorite wins. For this reason, the betting game tends to favor models that produce more extreme probability estimates towards the end of the game.

Table \ref{tab:betting-starts} and \ref{subtable Betting Game2} show the results of the betting game when bets are only made at the start of the game.  Again, the score-dependent Massey model outperforms all other models.  Interestingly, the null model shows a betting edge over every model except the score-dependent Massey model. Because the null model assigns each player a $50\%$ chance of winning at the start of the game, its positive profit indicates that the other models systematically overestimated the favorite’s probability of winning.

\begingroup
\setlength{\tabcolsep}{12pt}
\renewcommand{\arraystretch}{1.35} 
\begin{table}[htbp]
\centering
\begin{tabular}{|r|rrrrr|}
\hline
 & \multicolumn{5}{c|}{\normalsize Betting Model} \\
\hline
Odds Setting Model & Null & Logistic & Basic Sim. & Adjusted Sim. & SDMM \\
\hline
Null          & \color{lightgray} 0.000000  
              & \cellcolor{Dist1!27.886276}27.886276 
              & \cellcolor{Dist1!26.155935}26.155935 
              & \cellcolor{Dist1!43.078345}43.078345   
              & \cellcolor{Dist1!19.239773}19.239773\\

Logistic      & \cellcolor{Dist1!46.891051}46.891051 
              & \color{lightgray} 0.000000  
              & \cellcolor{Dist1!50.190971}50.190971
              & \cellcolor{Dist1!62.091096}62.091096  
              & \cellcolor{Dist1!42.724151}42.724151\\

Basic Sim.    & \cellcolor{Dist1!12.971015}12.971015 
              & \cellcolor{Dist1!18.001276}18.001276  
              & \color{lightgray} 0.000000 
              & \cellcolor{Dist1!8.932073}8.932073  
              & \cellcolor{Dist1!16.188947}16.188947\\

Adjusted Sim. & \cellcolor{Dist1!31.809787}31.809787 
              & \cellcolor{Dist1!31.817762}31.817762 
              & \cellcolor{Dist1!10.848435}10.848435 
              & \color{lightgray} 0.000000  
              & \cellcolor{Dist1!31.598307}31.598307\\

SDMM          & \cellcolor{DistMISS!14}-4.356340
              & \cellcolor{Dist1!2}0.123262 
              & \cellcolor{Dist1!5.777753}5.777753 
              & \cellcolor{Dist1!19.270752}19.270752   
              & \color{lightgray} 0.000000\\
\hline
\end{tabular}

\caption{Net profits from the betting game when bets are placed at the start of each round.}
\label{tab:betting-all}
\end{table}

\begingroup
\setlength{\tabcolsep}{12pt}
\renewcommand{\arraystretch}{1.35} 
\begin{table}[htbp]
\centering
\begin{tabular}{@{\extracolsep{\fill}}|r|rrrrr|}
\hline
 & \multicolumn{5}{c|}{\normalsize Betting Model} \\
\hline
Odds Setting Model & Null & Logistic & Basic Sim. & Adjusted Sim. & SDMM \\
\hline
Null
  & \color{lightgray}0.000000
  & \cellcolor{Dist1!21}5.332264
  & \cellcolor{Dist1!23}5.747500
  & \cellcolor{Dist1!47}10.299500
  & \cellcolor{Dist1!13}3.544219 \\

Logistic
  & \cellcolor{Dist1!35}7.893946
  & \color{lightgray}0.000000
  & \cellcolor{Dist1!44}9.637539
  & \cellcolor{Dist1!62}12.948526
  & \cellcolor{Dist1!33}7.473478 \\

Basic Sim.
  & \cellcolor{Dist1!11}3.018817
  & \cellcolor{Dist1!18}4.347174
  & \color{lightgray}0.000000
  & \cellcolor{Dist1!7}2.483589
  & \cellcolor{Dist1!12}3.294958 \\

Adjusted Sim.
  & \cellcolor{Dist1!37}8.225139
  & \cellcolor{Dist1!38}8.312484
  & \cellcolor{Dist1!11}3.137911
  & \color{lightgray}0.000000
  & \cellcolor{Dist1!34}7.667917 \\

SDMM
  & \cellcolor{DistMISS!6}-1.074808
  & \cellcolor{Dist1!2}0.292769
  & \cellcolor{Dist1!5}1.404614
  & \cellcolor{Dist1!21}5.123251
  & \color{lightgray}0.000000 \\
\hline
\end{tabular}
\caption{Net profits from the betting game when bets are placed at the start of each game.}
\label{tab:betting-starts}
\end{table}
\endgroup

\begin{table}[htbp]
\centering
\begin{subtable}{0.47\linewidth}
\centering
\setlength{\tabcolsep}{6pt}
\renewcommand{\arraystretch}{1.35}
\begin{tabular}{|r|r|c|}
\hline
Superior Model & Inferior Model & Difference  \\
\hline
SDMM          & Logistic      & 42.600889 \\
Basic Sim.    & Logistic      & 32.189695 \\
Adjusted Sim. & Logistic      & 30.273334 \\
SDMM          & Null          & 23.596113 \\
Null          & Logistic      & 19.004775 \\
Basic Sim.    & Null          & 13.184920 \\
SDMM          & Adjusted Sim. & 12.327555 \\
Adjusted Sim. & Null          & 11.268558 \\
SDMM          & Basic Sim.    & 10.411194 \\
Basic Sim.    & Adjusted Sim. & 1.916362 \\
\hline
\end{tabular}
\caption{\centering Head-to-head comparison in the betting game when bets are placed at the start of each round.}
\label{subtable Betting Game}
\end{subtable}
\hfill
\begin{subtable}{0.47\linewidth}
\centering
\setlength{\tabcolsep}{6pt}
\renewcommand{\arraystretch}{1.35}
\begin{tabular}{|r|r|c|}
\hline
Superior Model & Inferior Model & Difference  \\
\hline
SDMM          & Logistic      & 7.180709 \\
Basic Sim.    & Logistic      & 5.290365 \\
Adjusted Sim. & Logistic      & 4.636042 \\
SDMM          & Null          & 4.619027 \\
Basic Sim.    & Null          & 2.728683 \\
Null          & Logistic      & 2.561682 \\
SDMM          & Adjusted Sim. & 2.544666 \\
Adjusted Sim. & Null          & 2.074361 \\
SDMM          & Basic Sim.    & 1.890344 \\
Basic Sim.    & Adjusted Sim. & 0.654322 \\
\hline
\end{tabular}
\caption{\centering Head-to-head comparison in the betting game when bets are placed only at the start of each game.}
\label{subtable Betting Game2}
\end{subtable}

\caption{Head-to-head comparisons in the betting game.}
\label{tab:betting-both}
\end{table}

\section{Conclusion} \label{Sect: Conclusion}
%In this study, we evaluated five \st{predictive} \lthought{mathematical} models \lthought{that predict} \st{for predicting} outcomes in Darts 271, using real gameplay data from the Roanoke College Minton Invitational. Despite the inherent randomness of darts and the variability of amateur play, our results show that mathematical modeling can meaningfully capture player strategies and improve predictive accuracy over simple baselines models. Among the models tested, SDMM achieved the strongest performance in terms of Brier Score, while the Adjusted Sim. performed best in the competitive betting-game evaluation. 

In this study, we evaluated five mathematical models that predict outcomes of Darts 271 games, using real game play data collected during the 2025 Roanoke College Minton Invitational Darts 271 tournament. Despite the inherent randomness of darts and the variability of amateur play, our results show that mathematical modeling can meaningfully capture player abilities and improve predictive accuracy over simple baseline models.  Among the five models tested, the score-dependent Massey model achieved the strongest performance in terms of both the Brier score and the head-to-head betting-game evaluation. In contrast, the logistic regression model performed the worst under both metrics. Based on the results of the betting game, we conclude that the probabilities generated by the score-dependent Massey model most accurately reflect reality, although the model may systematically err on the conservative side.

Beyond model performance, this project highlights the value of the Minton Invitational as a community-building effort and a source of authentic research opportunities. The tournament brings together students, faculty, and staff in a shared environment that supports both competition and fun.  

%Ultimately, the success of these models highlights the idea that even in games shaped by luck, mathematics offers powerful tools for competitive outcomes. Furthermore, we show how our research can bridge mathematics theory and real-world uncertainty in meaningful ways.

\subsection{Future Work}

A variation of the score-dependent Massey model can be applied to other sports.  Instead of weighting the two ratings by the current score, the ratings can be weighted by how far into the season the game occurs.  For example, suppose team A defeats team B in an NHL game, and this game is team A’s \nth{30} game of the season and team B’s \nth{32}.
Then the corresponding equation included in the linear system is \[\frac{52}{82}r_A^{(1)} + \frac{30}{82}r_A^{(2)} - \frac{50}{82}r_B^{(1)} - \frac{32}{82}r_B^{(2)} = 1.\]
Here the denominator 
82 reflects the length of the NHL regular season.
Each team carries two ratings; however, in this setting the second rating $r^{(2)}$ captures how strong the team becomes over the course of the season.  This second rating can then be used at season’s end to rank teams and predict which team is most likely to hoist Lord Stanley’s Cup.

Similarly, this the score-dependent Massey model can be modified to predict the winner of a race at any point during the competition.  In this context, a linear equation is added each time the leader passes a mile marker (or any chosen checkpoint).
For example, suppose racer A leads racer B during a 10-mile race.
If racer A reaches mile 3 while racer B is at mile 2.7, then the corresponding equation added to the system is 
\[\frac{7}{10}r_A^{(1)} + \frac{3}{10}r_A^{(2)} - \frac{7.3}{10}r_B^{(1)} - \frac{2.7}{10}r_B^{(2)} = \pm1,\]
where the sign on the right-hand side is positive if A ultimately wins the race and negative if B does.

The score-dependent Massey model can contribute not only to sports analytics, but could also be repurposed as an experimental comparison tool between two or more technological products. For example, if one wanted to compare the performance of several food delivery robots, an experiment could be conducted where the researchers record the location of each robot when the leader reaches certain landmarks along their path. These observations would generate a system of equations analogous to the one described for racing competitions.
Because the robots’ performance contains inherent randomness, results will vary from trial to trial. By training the score-dependent Massey model on the collected dataset, researchers can obtain performance ratings for each robot and use these ratings to predict which robot is most likely to deliver food the fastest in future scenarios.

\subsection{Acknowledgments}
We would like to thank the Roanoke College Summer Scholars program for sponsoring this research project and the Roanoke College Information Technology department for collaborating with us to develop the data collection website.

\printbibliography
\end{document}